\documentclass[reprint,aps,pra,
longbibliography,
nofootinbib
]{revtex4-1}
\usepackage{graphicx}
\usepackage{amsmath,amssymb,mathrsfs,amsthm}
\usepackage[pdftex, colorlinks=true, linkcolor=blue, citecolor=blue, urlcolor=blue]{hyperref}
\usepackage{bbold}
\usepackage{comment}
\usepackage{units}

\usepackage{color}

\newcommand{\R}{\mathbf{R}}
\newcommand{\Rs}{\mathbf{R}_{s}}
\newcommand{\ds}{\mathbf{d}_{s}}
\newcommand{\q}{\mathbf{q}}
\newcommand{\ta}{\mathbf{t}_{1}}
\newcommand{\tb}{\mathbf{t}_{2}}
\newcommand{\Hpl}{H_{\mathrm{pl}}}
\newcommand{\Heh}{H_{\mathrm{eh}}}
\newcommand{\Hph}{H_{\mathrm{ph}}}
\newcommand{\Hpleh}{H_{\mathrm{pl\textrm{-}eh}}}
\newcommand{\Hplph}{H_{\mathrm{pl\textrm{-}ph}}}
\newcommand{\bsig}{b_{s}^{\sigma}}
\newcommand{\bsigdag}{{b_{s}^{\sigma}}^{\dagger}}
\newcommand{\fsig}{f_{ss'}^{\sigma\sigma'}(\q)}
\newcommand{\rhos}{\boldsymbol{\rho}_{ss'}}
\newcommand{\betatau}{\beta_{\tau}^{\varepsilon}}
\newcommand{\betataudag}{{\beta_{\tau}^{\varepsilon}}^{\dagger}}
\newcommand{\betataup}{\beta_{\tau'}^{\varepsilon'}}
\newcommand{\betataudagp}{{\beta_{\tau'}^{\varepsilon'}}^{\dagger}}
\newcommand{\ucoeff}{u_{\tau s}^{\varepsilon\sigma}(\q)}
\newcommand{\vcoeff}{v_{\tau s}^{\varepsilon\sigma}(\q)}
\newcommand{\omegatau}{\omega_{\tau}^{\varepsilon}(\q)}

\newcommand{\kp}{\mathbf{k}}
\newcommand{\lambdak}{\hat{\lambda}_{\kp}}
\newcommand{\omegak}{\omega_{\kp}}

\begin{document}

\title{Plasmons in two-dimensional lattices of near-field coupled nanoparticles}

\author{Fran\c cois Fernique}
\affiliation{Universit\'e de Strasbourg, CNRS, Institut de Physique et Chimie des Mat\'eriaux de Strasbourg, UMR 7504, F-67000 Strasbourg, France}
\author{Guillaume Weick}
\email{guillaume.weick@ipcms.unistra.fr}
\affiliation{Universit\'e de Strasbourg, CNRS, Institut de Physique et Chimie des Mat\'eriaux de Strasbourg, UMR 7504, F-67000 Strasbourg, France}

\begin{abstract}
We consider plasmonic metasurfaces constituted by an arbitrary periodic arrangement of spherical metallic nanoparticles. Each nanoparticle supports three degenerate dipolar localized surface plasmon (LSP) resonances. In the regime where the interparticle distance is much smaller than the optical or near-infrared wavelength associated with the LSPs, the latter couple through the dipole--dipole interaction and form collective plasmonic modes which extend over the whole metasurface. Within a Hamiltonian model which we solve exactly, we derive general expressions which enable us to extract analytically the quasistatic plasmonic dispersion for collective modes polarized within the plane and perpendicular to the plane of the metasurface. Importantly, our approach allows us not only to consider arbitrary Bravais lattices, but also non-Bravais two-dimensional metacrystals featuring nontrivial topological properties, such as, e.g., the honeycomb or Lieb lattices. 
Additionally, using an open quantum system approach, we consider perturbatively the coupling of the collective plasmons to both photonic and particle-hole environments, which lead, respectively, to radiative and nonradiative frequency shifts and damping rates, for which we provide closed-form expressions. 
The radiative frequency shift, when added to the quasistatic dispersion relation, provides an approximate analytical description 
of the fully retarded band structure of the collective plasmons. 
While it is tempting to make a direct analogy between the various systems which we consider and their electronic tight-binding equivalents, we critically examine how the long-range retarded and anisotropic nature of the dipole--dipole interaction may quantitatively and qualitatively modify the underlying band structures and  
discuss their experimental observability.
\end{abstract}

\maketitle

\section{Introduction}

The interaction of light with a small metallic particle results in collective electronic modes termed localized surface plasmons (LSPs) \cite{Bertsch1994, Kreibig1995}. In the case where the wavelength of the incoming light is much larger than the nanostructure itself, the LSP corresponds to a dipolar collective oscillation of the electronic cloud against the inert ionic background. While such a phenomena was empirically discovered centuries ago by late Romans~\cite{frees07}, the underlying physics was only theorized by Mie at the beginning of the 20th century, who solved Maxwell's equations for a metallic sphere embedded in a dielectric medium \cite{Mie1908, Born1980, Bohren2004}. Since then, 
plasmonic nanostructures have attracted a surge of interest due to their ability to perform subwavelength optics  by confining the electromagnetic field to nanometric regions \cite{Barnes-Nature2003, Maier-2007, Gramotnev-Natphot2010}. 
Due to the extreme sensitivity of the LSP resonance frequency to the nanoparticle size, shape, material, and dielectric environment \cite{Kreibig1995, Kelly2003}, a wealth of appealing technological applications have risen from the field of nanoplasmonics, such as, e.g., chemical \cite{Mayer2011} and biological~\cite{Anker2008} sensors.

When two metallic nanoparticles are positioned in close vicinity of each other (i.e., separated by a distance typically smaller than the LSP wavelength) so that they form a dimer, another factor crucially influencing the resonance frequencies of the latter is the Coulomb interaction between the LSPs. The quasistatic dipole--dipole interaction, which decays with the interparticle distance $d$ as $1/d^3$, gives rise to coupled modes, often coined ``hybridized'' modes \cite{jain10_CPL}, which correspond to symmetric (in-phase) or antisymmetric (out-of-phase) configurations of the oscillating electric dipolar moments on each nanoparticle. For transverse-polarized modes (with respect to the axis formed by the dimer), the high- (low-) energy plasmonic state corresponds to an in-phase (out-of-phase) configuration. Conversely, for longitudinal modes, the low- (high-) energy state corresponds to aligned
(antialigned) dipole moments. The splitting in frequencies between these hybridized modes scales with the interparticle distance as $1/d^3$, and can be spectroscopically resolved as long as the linewidth (which is of both radiative and nonradiative nature) of the two resonance peaks is somewhat smaller than the above-mentioned splitting.
The picture above is valid as long as the two nanoparticles are not too close to each other, so that higher multipolar modes do not mix with the dipolar ones \cite{proda03_Science, proda04_JCP, nordl04_NL, Park-PRB2004}, and the quantum tunneling of electronic charges between the two particles can be disregarded, such that so-called charge transfer plasmons are irrelevant \cite{Esteban-NatureCommun2012, Savage-Nature2012, Scholl-NanoLett2013, Esteban-Faraday2015, rossi15_PRL, zhu16_NatCommun}. 
Since the pioneering work by Ruppin \cite{Ruppin-PRB1982}, who extended Mie's theory \cite{Mie1908, Born1980, Bohren2004} 
to two nearby metallic spheres embedded in a dielectric medium, 
hybridized plasmonic modes in nanoparticle dimers have been investigated in numerous experimental
\cite{jain10_CPL,Tamaru-APl2002,Rechberger-OptCommun2003,Olk-NanoLett2008,Chu-Nanolett2009,Koh-ACSNano2009,Barrow-NanoLett2014, Savage-Nature2012, Scholl-NanoLett2013}
and theoretical works~\cite{Gerardy-PRB1983,nordl04_NL,Dahmen-NanoLett2007,Bachelier-PRL2008,Zuloaga-NanoLett2009,Esteban-NatureCommun2012,Zhang-PRB2014,Brandstetter-PRB2015,Brandstetter-PRB2016,downi17a_preprint, Savage-Nature2012, Scholl-NanoLett2013, Esteban-Faraday2015, rossi15_PRL}. 

In periodic arrays of near-field coupled nanoparticles, the dipolar interaction between the LSPs leads to collective modes that are extended over the whole lattice. In one-dimensional (1D) chains of regularly spaced nanoparticles, such collective plasmons were extensively studied both at the theoretical 
\cite{Brongersma-PRB2000,Maier-PRB2003,Park-PRB2004,Weber-PRB2004,Citrin-Nanolett2004,Simovski-PRE2005,Citrin-OptLett2006,Koenderink-PRB2006,Markel-PRB2007,Fung-OptLett2007,Lee-PRA2012,Compaijen-PRB2013,delPino-PRL2014,Compaijen-OptExpress2015,Petrov-PRA2015,Brandstetter-PRB2016,downi18_JPCM,Compaijen-PRB2018} and experimental
\cite{Krenn-PRL1999,Maier-PRB2002,Maier-PRB2003,Koenderink-PRB2007,Crozier-OptExpress2007,Apuzzo-NanoLett2013,Barrow-NanoLett2014,Gur-ArXiv2017} levels, since these systems may serve as plasmonic waveguides 
where plasmon--photon hybrid modes (so-called plasmon polaritons) are laterally confined to subwavelength scales and can possibly propagate over macroscopic distances. The importance of retardation effects in the dipolar interaction, which become relevant for nanoparticles in the chain spaced by a distance of the order of the LSP wavelength, was put forward in Refs.~\cite{Weber-PRB2004,Citrin-Nanolett2004,Simovski-PRE2005,Citrin-OptLett2006,Koenderink-PRB2006,Markel-PRB2007,Fung-OptLett2007,Compaijen-OptExpress2015,Petrov-PRA2015,downi18_JPCM,Compaijen-PRB2018}. 
In particular, it was shown that retardation leads to a pronounced discontinuity in the dispersion relation of the collective plasmons polarized transversely to the chain for wave vectors corresponding to the intersection of the light cone with the quasistatic band structure. The longitudinal dispersion relation remains however continuous, but presents discontinuities in its derivative in the vicinity of the above-mentioned crossing. Similar effects also occur in related condensed matter systems, such as exciton-polaritons in quantum wires \cite{citri92_PRL}.
The crucial role played by radiative and absorption losses on the propagation of plasmonic waves along the nanoparticle chain was also studied in detail in the previous works of Refs.~\cite{Brongersma-PRB2000,Maier-PRB2003,Park-PRB2004,Weber-PRB2004,Citrin-Nanolett2004,Simovski-PRE2005,Citrin-OptLett2006,Koenderink-PRB2006,Markel-PRB2007,Fung-OptLett2007,Lee-PRA2012,Compaijen-PRB2013,delPino-PRL2014,Compaijen-OptExpress2015,Petrov-PRA2015,Compaijen-PRB2018,downi18_JPCM}. Notably, 
Ref.~\cite{Brandstetter-PRB2016} showed that the nonradiative Landau damping, that is, the disintegration of the collective plasmons into particle-hole pairs, is of primarily importance as it dominates the plasmon linewidth for nanoparticles of only a few nanometers in size. 

Recently, dimerized  \cite{Ling-OptExpress2015,Downing-PRB2017,Gomez-ACSPhoton2017, pococ18_ACSPhoton, downi18_EPJB, pococ19_preprint, downi19_preprint} as well as zig zag chains of nanoparticles \cite{Poddubny-ACSPhoton2014, Sinev-Nanoscale2015} were proposed as a plasmonic analog of the celebrated Su-Schrieffer-Heeger (SSH) model~\cite{Su1979, Su1980, Heeger1988} presenting nontrivial topologically protected edge states. In particular, 
the robustness of such topological states against the long range retarded dipolar interactions was discussed in Refs.~\cite{pococ18_ACSPhoton, downi18_EPJB, pococ19_preprint}. 

The extension of the concepts introduced above to two spatial dimensions offers new exciting possibilities. 
Metasurfaces, that is, two-dimensional (2D) periodic arrangements of subwavelength metallic nanostructures, 
are indeed at present a very active field of research, as they enable one to tailor light in a way that goes far beyond what can be achieved with conventional optics. Thus far, the vast majority of the litterature on plasmonic metasurfaces (see, e.g., the review articles of Refs.~\cite{Meinzer-Natphot2014, glybo16_PhysRep, krave18_ChemRev, cherq19_ACR} and references therein) focused on the regime where the separation distance between each resonant element is of the order of the LSP wavelength, as this can be experimentally achieved with nowadays nanofabrication techniques. In this regime, the diffractive electromagnetic far fields generated by the essentially noninteracting nanoparticles of the array interfere and give rise to so-called surface lattice resonances (SLRs). 
The latter are of particular interest since they lead to much narrower absorption lines as that of the individual constituents of the metasurface, 
as well as angle-dependent dispersions, as was theoretically predicted in 
Refs.~\cite{carro86_JOSAB, zou04_JCP} and later experimentally verified in Refs.~\cite{Kravets-PRL2008, Auguie-PRL2008, chu08_APL}. Further works have demonstrated the use of SLRs in tailoring frequency stop gaps~\cite{rodri11_PRX} and are of relevance to applications in light emission \cite{rodri12_APL, Giannini-PRL2010}. 
Genuinely quantum-mechanical effects \cite{tame13_NaturePhys}, such as the exciting perspective of lasing~\cite{zhou13_NatureNanotech, Schokker-PRB2014, hakal17_NatureComm, guo19_PRL}, as well as Bose--Einstein 
condensation \cite{marti14_PRA, hakal18_NaturePhys}, have also been demonstrated in 2D plasmonic lattices. Notably, the works of Refs.~\cite{Humphrey-PRB2014, Guo-PRB2017} combining modeling and experiments have brought attention to the role of the geometrical arrangement of the nanostructures composing the metasurface on the SLR properties.

In the present work, we focus on the less explored case of near-field coupled nanoparticles supporting dipolar LSPs in metasurfaces. In this case, the stronger dipolar coupling between LSPs, as compared to the weak diffractive couplings encountered in SLRs, can exhibit potentially interesting analogies with atomically thin, 2D materials, such as graphene \cite{RevPhys2009} or transition metal dichalcogenides~\cite{mak16_NaturePhoton}, where the electronic band structures are usually well described by tight-binding calculations. Metasurfaces composed of near-field nanoparticles
may indeed present appealing nontrivial features in their band structure, paving the way to topological photonics performed with subwavelength elements \cite{lu14_NaturePhoton, ozawa19_RMP, rider19_JAP}.
For instance, it was theoretically demonstrated that a honeycomb lattice of plasmonic nanoparticles that are near-field coupled present chiral massless Dirac-like bosonic collective excitations~\cite{Han-PRL2009, Weick-PRL2013, Sturges-2DMat2015} which behave as electrons in 
graphene~\cite{RevPhys2009}. Such a honeycomb lattice further hosts topologically protected edge states~\cite{Wang_2016}. Exotic, so-called type-II Dirac plasmon-polaritons presenting a fully tunable tilted conical dispersion, were also recently unveiled in Ref.~\cite{Mann-NatCom2018}. Straining the honeycomb
metasurface and modifying the electromagnetic environment was recently shown to induce tunable
pseudo-magnetic fields for polaritons \cite{mann20_preprint}.

Due to the vast number of possible 2D lattices of near-field coupled plasmonic nanoparticles with potentially interesting properties in their band structure, here we develop a general theoretical framework which enables us to consider the plasmonic properties of
arbitrary metasurfaces. Our open quantum system approach, which builds on previous works on plasmonic 
dimers \cite{Brandstetter-PRB2015, downi17a_preprint} and chains~\cite{Brandstetter-PRB2016, downi18_JPCM}, allows us 
to unveil analytical expressions for the quasistatic plasmonic dispersion relations for collective modes polarized parallel and perpendicular to the plane of the metasurface. 
By considering the coupling of the purely plasmonic modes to photons of the electromagnetic vacuum, we also consider perturbatively the effects of retardation in the light--matter 
interaction, and we show that such retardation effects play a crucial role on the plasmonic band structure. Our approach, that straightforwardly include Ohmic losses, further gives access to the
radiative lifetime of the plasmonic modes, which we evaluate analytically. Importantly, we also consider the decay of the collective plasmons into electron-hole pairs and show that the resulting Landau damping can be as significant as it is in single nanoparticles which are only a few nanometers in size. 

The analytical treatment of the light--matter interaction has been shown  \cite{downi18_JPCM} in the case of chains to provide a rather good
account of the results stemming from laborious numerical calculations based on fully retarded solutions to Maxwell's equations for both the plasmonic dispersion relation and the radiative linewidth.
In particular, the discontinuity presented by the dispersion relation (by its derivative) for the transverse (longitudinal) modes is qualitatively
captured by the perturbative open quantum system approach.  
When we compare the results of the present study for 2D arrays of nanoparticles to existing numerical classical calculations 
for the square~\cite{zhen08_PRB} and the honeycomb lattices \cite{Han-PRL2009}, we further obtain good agreement.

Our model is not only applicable to plasmonic nanoparticles, but is also of relevance for any 2D system of regularly spaced subwavelength resonators coupled through the dipole--dipole interaction within the near-field regime, such as, e.g., dielectric nanoparticles where Ohmic losses are notably reduced 
\cite{Du2011, Slobo2015, Bakker2017}, microwave resonators \cite{Mann-NatCom2018}, magnonic microspheres \cite{Pirm2018}, cold atoms~\cite{Perczel2017}, or supercrystals made of semiconducting quantum dots~\cite{baimu13_SciRep}.  

This paper is organized as follows: Section \ref{sec.model} presents our Hamiltonian model and the open quantum system approach which we use to study collective plasmons in a generic 2D array of interacting spherical metallic nanoparticles. 
In Sec.~\ref{sec:QSB} we present the diagonalization procedure of the purely plasmonic Hamiltonian, which gives access to the quasistatic dispersion relation of the collective modes. The latter is then 
extensively discussed, including the cases of Bravais and non-Bravais lattices. 
In Secs.~\ref{sec:rad_frequency_shifts} and \ref{sec:rad_linewidths}, we consider the effects of the photonic environment alone which we treat perturbatively and present our results for the dispersion relation of the plasmonic modes including retardation effects (Sec.~\ref{sec:rad_frequency_shifts}), as well as the corresponding radiative lifetimes (Sec.~\ref{sec:rad_linewidths}). Section \ref{sec:Landau} then analyzes the (possibly crucial) role played by the electronic environment onto the collective modes and presents our results for their Landau damping decay rates, as well as their associated electronic frequency shifts. In Sec.~\ref{sec:exp}, we discuss the experimental observability of the plasmonic modes, before we conclude in Sec.~\ref{sec:conclusion}.

\section{Model}
\label{sec.model}

\begin{figure}
\begin{center}
\includegraphics[width=.8\columnwidth]{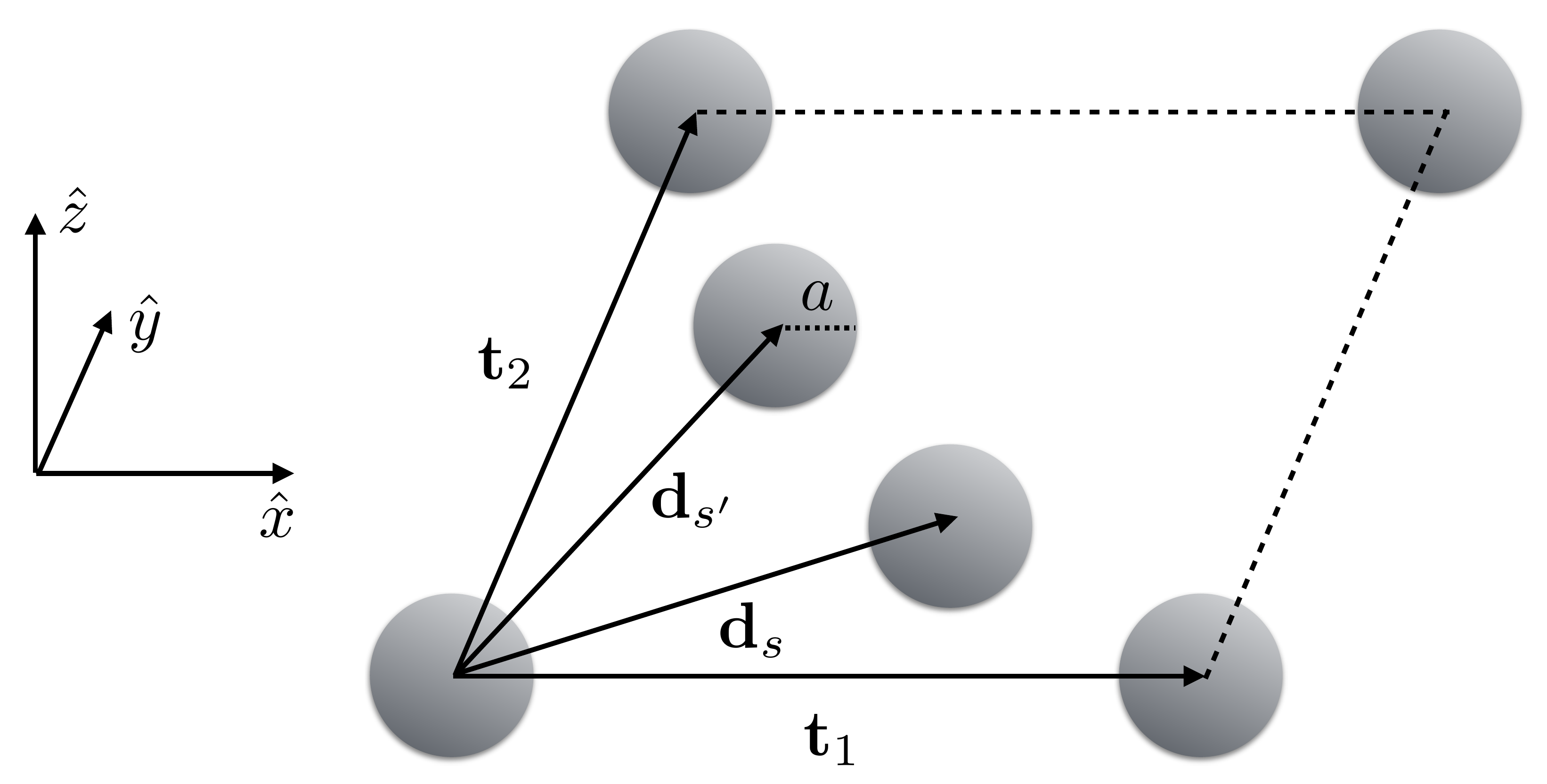}
\caption{\label{Lattices-examples} Sketch of the primitive cell of a generic 2D array of identical spherical metallic nanoparticles with radius $a$ forming a Bravais lattice with a basis. In the figure, $\mathbf{t}_{1}$ and $\tb$ are the primitive vectors of the Bravais lattice and $\ds$ ($s=1,\ldots,\mathcal{S}$) are the vectors forming the basis. By convention, $\mathbf{d}_{1}=0$.}
\end{center}
\end{figure}

We consider an ensemble of interacting spherical metallic, nonmagnetic nanoparticles of radius $a$ forming an arbitrary 2D Bravais lattice with a basis. The array is characterized by the vectors $\R=n\mathbf{t}_{1}+m\mathbf{t}_{2}$ forming the Bravais lattice. Here, $\mathbf{t}_{1}$ and $\mathbf{t}_{2}$ are the primitive lattice vectors, while $n\in[0, \mathcal{N}_{1}]$ and $m\in[0, \mathcal{N}_{2}]$ are integer numbers with $\mathcal{N}_{1}$ ($\mathcal{N}_{2}$) the number of unit cells in the $\mathbf{t}_{1}$ ($\tb$) direction. The array is composed of $\mathcal{S}$ sublattices, and the nanoparticles belonging to the sublattice $s=1,\ldots,\mathcal{S}$ are located at $\Rs=\R+\ds,$ where $\ds$ is the vector belonging to the $xy$ plane and connecting the sublattice $s$ to $\R$ (see Fig.\ \ref{Lattices-examples}). By convention, we set $\mathbf{d}_{1}=0$ in the remainder of the paper.

Each nanoparticle forming the array supports three degenerate orthogonal dipolar LSP resonances polarized along the $\sigma=x$, $y$, and $z$ directions and characterized by the frequency $\omega_{0}$. Neglecting the effect of the embedding medium, as well as the screening of the valence electrons by the core electrons, one has $\omega_{0}=\omega_{\mathrm{p}}/\sqrt{3}$ \cite{Kreibig1995}, where $\omega_{\mathrm{p}}=\sqrt{4\pi n_{\mathrm{e}}e^{2}/m_{\mathrm{e}}}$ is the plasma frequency (throughout this paper, we use cgs units). Here, $-e$ ($<0$) is the electron charge, $m_{\mathrm{e}}$ its mass, and $n_{\mathrm{e}}$ the electronic density of the metal. 

The dipolar LSPs supported by the nanoparticles in the array interact with their neighbors through the Coulomb interaction. 
Such a coupling gives rise to collective plasmonic modes which extend over the whole metasurface~\cite{Meinzer-Natphot2014}. 
In this work we focus on the subwavelength regime where $d \ll k_{0}^{-1}$, with $d$ the center-to-center nearest neighbor interparticle distance and $k_{0}=\omega_{0}/c$, with $c$ the speed of light in vacuum. 
Furthermore, we assume that the interparticle distance $d\gtrsim 3a$.
In this parameter regime, Park and Stroud  \cite{Park-PRB2004} have shown by means of an exact diagonalization procedure based on a generalized tight-binding approach including multipoles up to $\ell=80$ (with $\ell$ the angular momentum) that the dominant interaction between two nanoparticles is the near-field quasistatic electric dipole--dipole ($\ell=1$) interaction which decays as $1/d^{3}$. For smaller interparticle distances ($2a\leqslant d \lesssim 3a$), higher multipolar interactions with $\ell>1$ \cite {Park-PRB2004} as well as tunneling phenomena \cite{zhu16_NatCommun} play an important role for the description of the collective modes, and we do not enter this regime of parameters in the remainder of the paper. Notice also that, since we consider nonmagnetic particles, and since our model does not incorporate an external magnetic field, magnetoplasmonic modes (see, e.g., Ref.~\cite{Weick-PRB2011}) which would result in oscillating magnetic dipolar moments are irrelevant to the present study. 

Similarly to individual LSPs, the extended plasmonic modes supported by the metasurface are coupled to both a photonic and an electronic environment. 
The collective plasmons are indeed coupled to vacuum electromagnetic modes through the light--matter interaction, giving rise to finite radiative lifetimes as well as radiative frequency shifts, stemming from the retarded part of the dipole--dipole interaction~\cite{downi17a_preprint,downi18_JPCM}.
Moreover, such collective modes are coupled to electron-hole excitations inside the nanoparticles, leading to the nonradiative Landau damping and an additional frequency renormalization. 

We write the full Hamiltonian of the system as
\begin{equation}
H=\Hpl+\Hph+\Heh+\Hplph+\Hpleh,
\label{System-Hamiltonian}
\end{equation}
where $\Hpl$ describes the purely plasmonic degrees of freedom, while $\Hph$ and $\Heh$ correspond to the photonic and electronic environments, respectively. In Eq.\ \eqref{System-Hamiltonian}, $\Hplph$ and $\Hpleh$ are the coupling Hamiltonians of the plasmonic subsystem to photons and electron-hole pairs, respectively.

Within the Coulomb gauge \cite{Cohen-1992, craig}, the plasmonic Hamiltonian 
\begin{equation}
\label{eq:H_pl}
\Hpl=\Hpl^{0}+\Hpl^{\mathrm{int}}
\end{equation}
describing the near-field coupled LSPs is characterized by a noninteracting and an interacting term \cite{Weick-PRL2013, Brandstetter-PRB2016}. The noninteracting part related to individual nanoparticles reads 
\begin{align}
H_{\mathrm{pl}}^{0}=\sum_{s}\sum_{\Rs}\sum_{\sigma}\left\{\frac{\left[\Pi_{s}^{\sigma}(\Rs)\right]^{2}}{2N_\mathrm{e}m_\mathrm{e}}+\frac{N_\mathrm{e}m_\mathrm{e}}{2}\omega_{0}^{2}\left[h_{s}^{\sigma}(\Rs)\right]^{2}\right\},
\label{equ.Hpl-0}
\end{align}
with $h_{s}^{\sigma}(\Rs)$ the $\sigma$ component of the displacement field associated with the dipole moment $\mathbf{p}(\mathbf{R}_s)=-eN_{\mathrm{e}}\sum_{\sigma}h_{s}^{\sigma}(\Rs)\hat\sigma$ of a single LSP located at position $\Rs$, while $\Pi_{s}^{\sigma}(\Rs)$ is the momentum conjugated to $h_{s}^{\sigma}(\Rs)$. 
Here and in what follows, hats designate unit vectors. 
In Eq.\ \eqref{equ.Hpl-0}, $N_{\mathrm{e}}$ is the total number of valence electrons in each nanoparticle. The interacting term in Eq.\ \eqref{eq:H_pl} arises from the quasistatic dipole--dipole interaction and reads
\begin{align}
H_{\mathrm{pl}}^{\mathrm{int}}=&\;\frac{(eN_{\mathrm{e}})^{2}}{2}\sum_{ss'}\sum_{\substack{\Rs \mathbf{R}'_{s'}\\ (\Rs\neq \mathbf{R}'_{s'})}}\sum_{\sigma\sigma'}
h_{s}^{\sigma}(\Rs)h_{s'}^{\sigma'}(\mathbf{R}'_{s'})\nonumber\\
&\times \frac{\delta_{\sigma\sigma'}-3(\hat{{\sigma}}\cdot\hat{{\rho}}_{ss'})(\hat{{\sigma}}'\cdot\hat{{\rho}}_{ss'})}{\left|\Rs-\mathbf{R}'_{s'}\right|^3}, 
\label{equ.Hpl-int}
\end{align}
where $\boldsymbol{\rho}_{ss'}=\mathbf{R}'_{s'}-\mathbf{R}_s$. 

As we deal with nanoparticles of small sizes, quantum-size effects (such as Landau damping) can be important. 
Moreover, a quantum treatment of the plasmonic degrees of freedom provides a self-contained description of the light--matter interaction \cite{downi17a_preprint, downi18_JPCM}. 
In view of the analysis of these effects, we hence write the  
plasmonic Hamiltonian \eqref{eq:H_pl}  in terms of 
the bosonic ladder operators
\begin{equation}
\label{eq:quantization}
b_{s}^{\sigma}(\Rs)=\sqrt{\frac{N_\mathrm{e}m_\mathrm{e}\omega_{0}}{2\hbar}}h_{s}^{\sigma}(\Rs)+\mathrm{i}\frac{\Pi_{s}^{\sigma}(\Rs)}{\sqrt{2N_\mathrm{e}m_\mathrm{e}\hbar\omega_{0}}}
\end{equation}
that annihilate an LSP at position $\Rs$ on sublattice $s$ with polarization $\sigma=x,y,z$
and its adjoint ${b_{s}^{\sigma}}^{\dagger}(\Rs)$. Together with Eqs.\ \eqref{equ.Hpl-0} and \eqref{equ.Hpl-int}, Eq.~\eqref{eq:H_pl} thus takes the form
\begin{align}
\label{eq:H_compa}
H_{\mathrm{pl}}=&\;\hbar\omega_{0}\sum_{s}\sum_{\Rs}\sum_{\sigma}\bsigdag(\Rs)\bsig(\Rs)
\nonumber\\
&+\frac{\hbar\Omega}{2}\sum_{ss'}\sum_{\substack{\Rs \mathbf{R}'_{s'}\\ (\Rs\neq \mathbf{R}'_{s'})}}\sum_{\sigma\sigma'}\left[\bsig(\Rs)+\bsigdag(\Rs)\right]\nonumber\\
&\times \left[b_{s'}^{\sigma'}(\mathbf{R}'_{s'})+{b_{s'}^{\sigma'}}^\dagger(\mathbf{R}'_{s'})\right]\nonumber\\
&\times \frac{\delta_{\sigma\sigma'}-3(\hat{{\sigma}}\cdot\hat{{\rho}}_{ss'})(\hat{{\sigma}}'\cdot\hat{{\rho}}_{ss'})}{\left(|\Rs-\mathbf{R}'_{s'}|/d\right)^3},
\end{align}
with the coupling constant
\begin{equation}
\label{eq:Omega}
\Omega=\frac{\omega_0}{2}\left(\frac{a}{d}\right)^3.
\end{equation}
Note that $\Omega\ll\omega_0$ since we consider interparticle distances $d\geqslant 3a$.

The Hamiltonian \eqref{eq:H_compa} displays some similarities with a tight-binding Hamiltonian of an electronic 2D system~\cite{Grosso-2014}. The first term on the right-hand side of Eq.~\eqref{eq:H_compa} [$\propto \bsigdag(\Rs)b_{s}^{\sigma}(\Rs)$] corresponds to a fixed on-site energy, while the resonant terms [$\propto \bsigdag(\Rs)b_{s'}^{\sigma'}(\mathbf{R}'_{s'})$] in the second term describe the creation of an LSP at the lattice site $\Rs$ together with the destruction of an LSP at the lattice site $\mathbf{R}'_{s'}$, similary to a hopping term. 

There are however important differences between the plasmonic Hamiltonian \eqref{eq:H_compa} and an electronic tight-binding Hamiltonian. 
Firstly, plasmons correspond to bosonic excitations, which do not have a finite chemical potential.
Secondly, the dipole--dipole interaction responsible for the existence of the collective plasmons is quite different compared to the hopping amplitude in a tight-binding model~\cite{Grosso-2014}. On the one hand, the long-range dipolar interaction scales with $1/|\Rs-\mathbf{R}'_{s'}|^{3}$, whereas the hopping amplitude decreases exponentially with the distance. Thus, the dipolar interaction beyond the first neighbors can have important effects. On the other hand, the dipole--dipole interaction depends on the polarization of the excitations, contrarily to those in tight-binding models. Thirdly, there are additional nonresonant terms [$\propto \bsigdag(\Rs){b_{s'}^{\sigma'}}^{\dagger}(\mathbf{R'}_{s'})+\mathrm{h.c.}$] in Eq.~\eqref{eq:H_compa} which do not conserve the number of quasiparticles and play a crucial role for physical quantities depending on the plasmonic eigenstates, e.g., the collective mode damping rates \citep{Brandstetter-PRB2016}. How the above-mentioned differences may crucially affect the plasmonic band structure is extensively discussed in Sec.\ 
\ref{sec:QSB}.

The Hamiltonian \eqref{System-Hamiltonian} further describes the coupling of the collective plasmons to vacuum electromagnetic modes in a volume $\mathcal{V}$ described by the Hamiltonian 
\begin{equation}
\Hph=\sum_{\kp,\lambdak}\hbar\omegak {a_{\kp}^{\lambdak}}^{\dagger}{a_{\kp}^{\lambdak}},
\label{H-ph}
\end{equation}
where $a_{\kp}^{\lambdak}$ (${a_{\kp}^{\lambdak}}^\dagger$) annihilates (creates) a photon with wave vector $\kp$, transverse polarization $\lambdak$ (i.e., $\kp\cdot\lambdak=0$), and dispersion $\omegak=c|\kp|$. In the long-wavelength limit $|\kp|a\ll1$, the minimal-coupling Hamiltonian between plasmons and photons in Eq.\ \eqref{System-Hamiltonian} reads~\cite{Cohen-1992}
\begin{align}
\Hplph=&\;\sum_{s}\sum_{\Rs}\left[\frac{e}{m_{\mathrm{e}}c}\mathbf{\Pi}_{s}(\Rs)\cdot\mathbf{A}(\Rs)
\right.
\nonumber\\
&\left.+\frac{N_{\mathrm{e}}e^2}{2m_\mathrm{e}c^2}\mathbf{A}^2(\Rs)\right],
\label{H-plph}
\end{align}
with
\begin{align}
\mathbf{\Pi}_{s}(\Rs)=&\;\mathrm{i}\,\sqrt{\frac{N_{\mathrm{e}}m_{\mathrm{e}}\hbar\omega_{0}}{2}}
\sum_{\sigma}\hat\sigma
\left[\bsigdag(\Rs)-\bsig(\Rs)\right]
\label{Momentum}
\end{align}
and where
\begin{align}
\mathbf{A}(\Rs)=\sum_{\kp,\lambdak}\lambdak \sqrt{\frac{2\pi\hbar c^2}{\mathcal{V}\omegak}}\left({a_{\kp}^{\lambdak}}
\mathrm{e}^{\mathrm{i}\kp\cdot\Rs}+{a_{\kp}^{\lambdak}}^{\dagger}\mathrm{e}^{-\mathrm{i}\kp\cdot\Rs}
\right)
\label{Vector-potential}
\end{align}
is the vector potential evaluated at the nanoparticle centers.
Note that since we consider interparticle separation distances much smaller than the wavelength associated with the LSP resonances, we neglect Umklapp processes in Eqs.~\eqref{H-ph} and \eqref{H-plph}.
Importantly, within the Coulomb gauge, the first term on the right-hand side of Eq.~\eqref{H-plph} contains the retardation effects stemming from the 
finite velocity of light \cite{craig}. 

In addition to the photonic environment, the collective plasmons are coupled to electron-hole excitations described by the Hamiltonian
\cite{Weick-PRB2005,Weick-PRB2006}
\begin{align}
\Heh=\sum_{s}\sum_{\Rs}\sum_{i}\epsilon_{\Rs i}c^{\dagger}_{\Rs i}c^{\phantom{\dagger}}_{\Rs i},
\label{H-eh}
\end{align}
where $c^{\phantom{\dagger}}_{\Rs i}$ ($c^{\dagger}_{\Rs i}$) annihilates (creates) an electron in the nanoparticle located at $\Rs$ associated with the one-body state $\vert\Rs i\rangle$ with energy $\epsilon_{\Rs i}$ in the self-consistent potential $V$ of that nanoparticle. Assuming $V$ to be a spherical hard-wall potential, the coupling between plasmonic and single electronic degrees of freedom is
\cite{Brandstetter-PRB2015}
\begin{align}
\Hpleh=&\;\Lambda\sum_{s}\sum_{\Rs}\sum_{\sigma}\sum_{ij}
\left[\bsig(\Rs)+\bsigdag(\Rs)\right]\nonumber\\
&\;\times\langle\Rs i\vert\;\sigma\;\vert\Rs j\rangle c^{\dagger}_{\Rs i}c^{\phantom{\dagger}}_{\Rs j},
\label{H-pleh}
\end{align}
with 
\begin{equation}
\label{eq:Lambda}
\Lambda=\sqrt{\frac{\hbar m_{\mathrm{e}}\omega_{0}^{3}}{2N_{\mathrm{e}}}}. 
\end{equation}

\section{Quasistatic plasmonic band structure}
\label{sec:QSB}
In this section, we start by focusing on the plasmonic degrees of freedom alone. We first present the diagonalization procedure of the Hamiltonian \eqref{eq:H_compa}, from which we obtain the quasistatic plasmonic band structure. The latter is then analyzed in detail for both the cases 
of Bravais (Sec.~\ref{sec:Bravais}) and non-Bravais lattices (Sec.~\ref{sec:omega_basis}). 
The discussion about how the band structure is influenced by retardation effects in the dipole--dipole interaction is postponed to 
Sec.~\ref{sec:rad_frequency_shifts}.

The  plasmonic Hamiltonian $\Hpl$ given in Eq.~\eqref{eq:H_compa} is quadratic, and can thus be diagonalized 
exactly by means of a bosonic Bogoliubov transformation. In the large metasurface limit, going to Fourier space, 
we then introduce a set of bosonic operators (see Appendix \ref{app:diag} for details)
\begin{align}
\betatau(\q)=\sum_{s\sigma}\left[\ucoeff\bsig(\q)+\vcoeff\bsigdag(-\q)\right]
\label{Beta-operators}
\end{align}
annihilating a collective plasmon with wave vector $\q$ in the band $\tau$ with polarization $\varepsilon$. Notice that, in general, $\varepsilon=\varepsilon_\tau(\q)$ is a $\q$- and $\tau$-dependent quantity but we drop in the remaining of this paper both indexes for notational simplicity. 
In Eq.~\eqref{Beta-operators}, 
$\bsig(\q)$ corresponds to the Fourier transform of the bosonic ladder operator \eqref{eq:quantization} [cf.\ Eq.~\eqref{eq:b_Fourier}], 
while $\ucoeff$ and $\vcoeff$ are complex coefficients which are determined by imposing that $\Hpl$ is diagonal in this new 
basis, i.e., 
\begin{align}
\Hpl=\sum_{\q}\sum_{\tau\varepsilon}H_\tau^\varepsilon(\q),
\quad
H_\tau^\varepsilon(\q)=\hbar\omegatau\betataudag(\q)\betatau(\q),
\label{Plasmonic-H-diagonal}
\end{align}
where $\omegatau$ is the quasistatic collective plasmon dispersion relation.
The Bogoliubov operators $\betatau(\q)$ and $\betataudag(\q)$ act on an eigenstate $|n_\tau^\varepsilon(\q)\rangle$ of the Hamiltonian $H_\tau^\varepsilon(\q)$ corresponding to a collective plasmon in the band $\tau$ with wave vector $\q$ and polarization $\varepsilon$ as
$\betatau(\q)|n_\tau^\varepsilon(\q)\rangle=\sqrt{n_\tau^\varepsilon(\q)}|n_\tau^\varepsilon(\q)-1\rangle$ and 
$\betataudag(\q)|n_\tau^\varepsilon(\q)\rangle=\sqrt{n_\tau^\varepsilon(\q)+1}|n_\tau^\varepsilon(\q)+1\rangle$, respectively. Here, $n_\tau^\varepsilon(\q)$ is a non-negative integer.

The dispersion relation $\omegatau$, as well as the coefficients of the Bogoliubov transformation \eqref{Beta-operators}, are obtained from the Heisenberg equation of motion [cf.\ Eq.~\eqref{eq:Heisenberg}], 
which yields the system of equations
\begin{subequations}
\label{Hopfield-coeff}
\begin{align}
\big[\omega_{0}&-\omegatau\big]\ucoeff\nonumber\\
&+\Omega\sum_{s'\sigma'}\big[u_{\tau s'}^{\varepsilon\sigma'}(\q)-v_{\tau s'}^{\varepsilon\sigma'}(\q)\big]f_{s's}^{\sigma'\sigma}(\q)=0
\label{Hopfield-u}
\end{align}
and
\begin{align}
-\big[\omega_{0}&+\omegatau\big]\vcoeff\nonumber\\
&+\Omega\sum_{s'\sigma'}\big[u_{\tau s'}^{\varepsilon\sigma'}(\q)-v_{\tau s'}^{\varepsilon\sigma'}(\q)\big]f_{s's}^{\sigma'\sigma}(\q)=0.
\label{Hopfield-v}
\end{align}
\end{subequations}

In Eq.\ \eqref{Hopfield-coeff} the lattice sum
\begin{align}
\fsig=&\;\sum_{\substack{\rhos\\ (\rho_{ss'}\neq0)}}\bigg(\frac{d}{\rho_{ss'}}\bigg)^{3}\mathrm{e}^{\mathrm{i}\q\cdot\rhos}\nonumber\\
&\times\left[\delta_{\sigma\sigma'}-3(\hat{{\sigma}}\cdot\hat{{\rho}}_{ss'})(\hat{{\sigma}}'\cdot\hat{{\rho}}_{ss'})\right]
\label{fsig}
\end{align}
takes into account the quasistatic dipolar interaction between each pair of nanoparticles composing the metasurface. In the remainder of the paper, the lattice sum \eqref{fsig} is calculated for a specific metasurface numerically, until satisfactory convergence is obtained. In practice, we perform the summation in Eq.~\eqref{fsig} up to $\rho_{ss'}^\mathrm{max}=300d$, 
which yields a relative error of the order of $10^{-9}$.

The system of equations \eqref{Hopfield-coeff} needs to be satisfied for all integer $s \in [1, \mathcal{S}]$ and for all polarizations $\sigma=x, y, z$, yielding a $6\mathcal{S} \times 6\mathcal{S}$ eigensystem. Due to the structure of the lattice sum \eqref{fsig}, such an eigensystem decouples into a block-diagonal matrix composed of a $4\mathcal{S}\times 4\mathcal{S}$ and a $2\mathcal{S}\times 2\mathcal{S}$ block, corresponding to the in-plane (IP, $\sigma=x,y$) and out-of-plane (OP, $\sigma=z$) polarized modes, respectively. Each block then yields a secular equation of order $2\mathcal{S}$ and $\mathcal{S}$ in $[\omegatau]^{2}$, respectively, which then gives access to the quasistatic band structure. 

Additional insight about the nature of the quasistatic collective plasmons can be obtained from their corresponding eigenstates, from which we can deduce the polarization $\varepsilon$ of the collective modes. Introducing the vector 
$\mathbf{u}_\tau^\varepsilon(\q)=(\sum_s u_{\tau s}^{\varepsilon x}(\q), \sum_s u_{\tau s}^{\varepsilon y}(\q), \sum_s u_{\tau s}^{\varepsilon z}(\q))$, we define the polarization angle as \cite{lamow18_PRB}
\begin{equation}
\label{eq:polarization}
\phi_\tau^\varepsilon(\q)=\arccos{\left(\left|\hat u_\tau^\varepsilon(\q)\cdot\hat{q}\right|\right)}. 
\end{equation}
With such a definition, purely longitudinal (transverse) collective plasmons correspond to 
$\phi_\tau^\varepsilon(\q)=0$ ($\pi/2$).

\subsection{Bravais lattices}
\label{sec:Bravais}
In the case of an arbitrary Bravais lattice (i.e., $\mathcal{S}=1$), the sublattice indexes $s$ and $s'$, as well as the band index $\tau$ are irrelevant and can then be dropped from the system of equations~\eqref{Hopfield-coeff}. Such a system can be fully analytically solved, yielding 
for the OP plasmonic modes polarized in the $\varepsilon=z$ direction the dispersion relation 
\begin{equation}
\label{equ.omega-zz-simple}
\omega^{z}(\q)=\omega_{0}\sqrt{1+2\frac{\Omega}{\omega_{0}}f^{zz}(\q)},
\end{equation}
and 
\begin{align}
\omega^{\varepsilon_{\parallel,\pm}}(\q)=&\;\omega_{0}\bigg\{1+\frac{\Omega}{\omega_{0}}\bigg[f^{xx}(\q)+f^{yy}(\q) \nonumber \\
&\pm\sqrt{\left[f^{xx}(\q)-f^{yy}(\q)\right]^{2}+4\left[f^{xy}(\q)\right]^{2}}\bigg]\bigg\}^{1/2}
\label{equ.omega-xy-simple}
\end{align}
for the IP modes with polarizations $\varepsilon=\varepsilon_{\parallel,\pm}$. Explicit expressions for the corresponding Bogoliubov coefficients 
can be found in Appendix \ref{app:diag}.

We notice that the quasistatic dispersion relations \eqref{equ.omega-zz-simple} and \eqref{equ.omega-xy-simple} have been obtained by Zhen \textit{et al.}~\cite{zhen08_PRB} using a classical point-dipole model [cf.\ their Eq.~(3.3)]. It is not surprising that, within the quasistatic approximation, our quantum approach and the classical one of Ref.~\cite{zhen08_PRB} (which however is limited to simple Bravais lattices) give the same results since the Hamiltonian at hand [cf.\ Eq.~\eqref{eq:H_pl}] is quadratic. 

\begin{figure}[t!p]
\begin{center}
\includegraphics[width=\columnwidth]{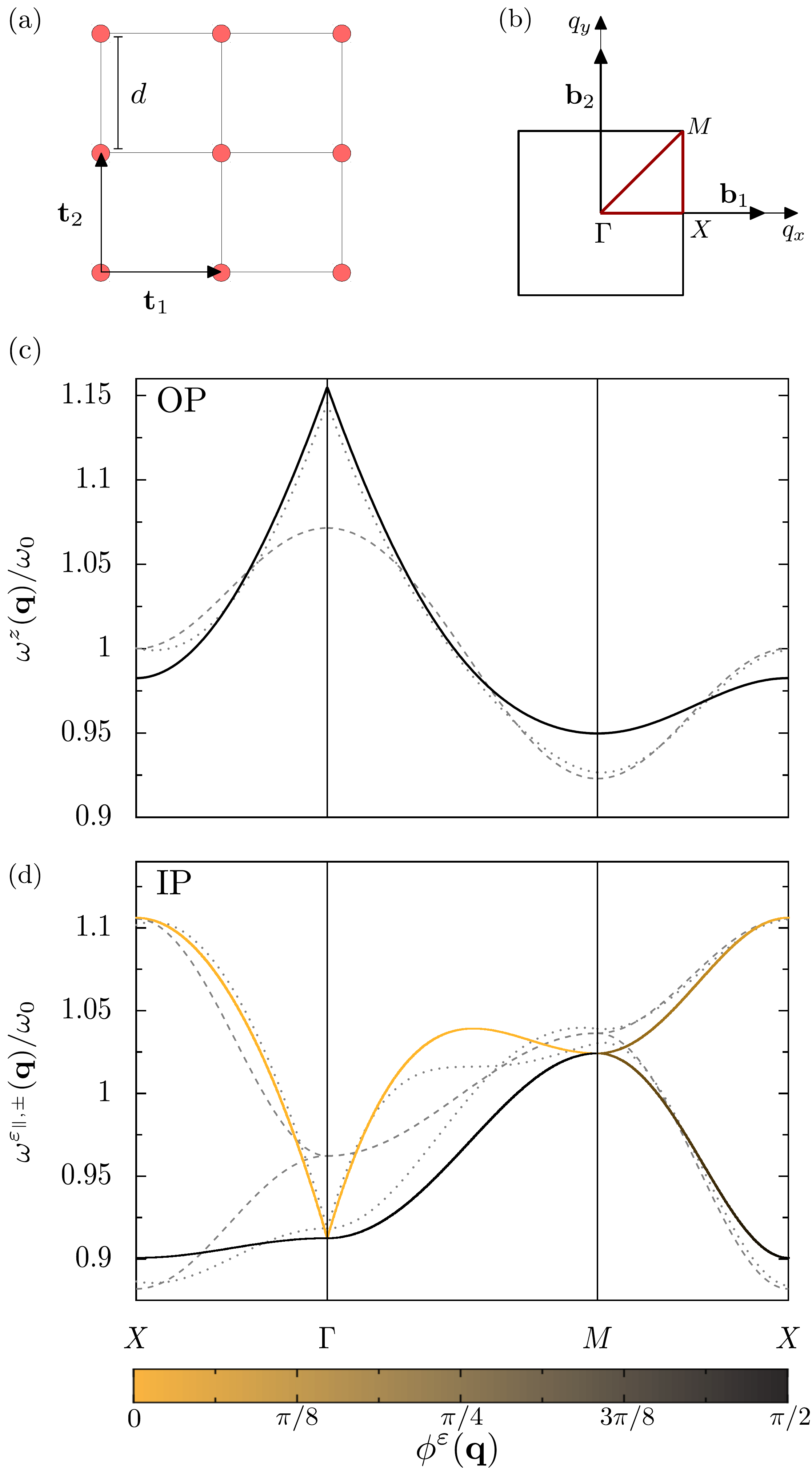}
\caption{\label{fig.Disp-square-z} (a) Sketch of a square lattice with primitive lattice vectors $\ta=d\, (1,0)$ and $\tb=d\, (0,1)$. (b) Corresponding first Brillouin zone, with primitive reciprocal vectors $\mathbf{b}_1=\frac{2\pi}{d}(1,0)$ and $\mathbf{b}_2=\frac{2\pi}{d}(0,1)$. (c) and (d) Quasistatic plasmonic dispersion relation as a function of the wave vector $\mathbf{q}$ (scaled with the interparticle distance $d$) along high-symmetry paths in the first Brillouin zone [cf.\ (b)] for (c) out-of-plane (OP) and (d) in-plane (IP) polarizations. The solid lines represent the full quasistatic dispersion, including long-range couplings, and the color code corresponds to the polarization angle \eqref{eq:polarization}, which equals $0$ ($\pi/2$) for purely longitudinal (transverse) modes. The dashed and dotted lines correspond to the nearest-neighbor and mean-field approximations discussed, respectively, in the main text and in Appendix \ref{app:MF}. In the figure, the interparticle distance $d=3a$ (corresponding to $\Omega=\omega_{0}/54$).}
\end{center}
\end{figure}

As an example of application of our general method for obtaining the quasistatic band structure of plasmonic modes in generic Bravais lattices, we consider in the following the simple square lattice sketched in Fig.\ \ref{fig.Disp-square-z}(a), whose corresponding first Brillouin zone (1BZ) is depicted in Fig.~\ref{fig.Disp-square-z}(b). 
The plasmonic dispersion relation \eqref{equ.omega-zz-simple} for the square lattice is plotted in Fig.~\ref{fig.Disp-square-z}(c) for the OP polarization as a solid line. For comparison, we also show (dashed line) the plasmonic band structure considering only dipolar interactions between nearest neighbors (nn) in the lattice, for which the lattice sum \eqref{fsig} reduces to $f_{\mathrm{nn}}^{zz}(\q)=2[\cos(q_{x}d)+\cos(q_{y}d)]$. As can be seen from Fig.\ \ref{fig.Disp-square-z}(c), the nearest-neighbor approximation qualitatively reproduces the full band structure in most of the 1BZ, except for wave numbers close to the $\Gamma$ point. There, the long-range nature of the quasistatic dipolar interaction leads to a pronounced cusp of the dispersion relation. Such a behavior, which is purely an artifact of the quasistatic approximation and which points to the importance of taking into account retardation effects, as we shall do in Sec.~\ref{sec:rad_frequency_shifts}, can be understood by a mean-field treatment of the dipolar interactions beyond those between nearest neighbors (see  Appendix \ref{app:MF}).

We now turn to the discussion of the plasmonic modes polarized within the plane formed by the square lattice. The band structure \eqref{equ.omega-xy-simple} is 
plotted in Fig.\ \ref{fig.Disp-square-z}(d) as solid lines. 
The color code corresponds to the polarization angle defined in Eq.~\eqref{eq:polarization}. While for the high-symmetry axes $\Gamma M$ or $\Gamma X$, the IP collective plasmons are purely longitudinal or transverse, for less-symmetric axes such as the $MX$ direction in the 1BZ, such modes can be of a mixed type. 
For comparison, we further plot the dispersion relation taking into account nearest-neighbor couplings only, 
for which the lattice sums in Eq.~\eqref{equ.omega-xy-simple} are replaced by
$f^{xx}_\mathrm{nn}(\q)=-4\cos{(q_xd)}+2\cos{(q_yd)}$, 
$f^{yy}_\mathrm{nn}(\q)=2\cos{(q_xd)}-4\cos{(q_yd)}$, and
$f^{xy}_\mathrm{nn}(\q)=0$. 
In contrast to the OP modes [Fig.~\ref{fig.Disp-square-z}(c)], the long-range nature of the dipolar interaction has a more pronounced effect on the plasmonic band structure for IP polarized modes. For instance, the dipolar interaction lifts the degeneracy of the dispersion induced by the symmetry of the square lattice  within the nearest-neighbor approximation along the $\Gamma M$ direction of the 1BZ. In addition, the long-range dipolar interaction leads to a cusp of the upper plasmonic band at the $\Gamma$ point, while the lower band does not show such a singularity in the derivative of $\omega^{\varepsilon_{\parallel,\pm}}(\q)$. This behavior can be explained along the lines of the mean-field approximation discussed in Appendix \ref{app:MF}.

\subsection{Bravais lattices with a basis}
\label{sec:omega_basis}
Our general method for obtaining quasistatic plasmonic band structures further applies to arbitrary Bravais lattices with a basis. 
In the following, we start by considering Bravais lattices with a basis of two.

\subsubsection{Bipartite lattices}
\label{sec:bipartite}
In the case of a bipartite lattice ($\mathcal{S}=2$), the $4\times4$ matrix resulting from the system of equations \eqref{Hopfield-coeff} for the OP 
polarization $\sigma=z$ can be straightforwardly solved, yielding the two bands with dispersion relations 
\begin{equation}
\label{eq:omega_z_bipartite}
\omega_\tau^z(\q)=\omega_0
\sqrt{1+2\frac{\Omega}{\omega_0}\big[f_{11}^{zz}(\q)+\tau |f_{12}^{zz}(\q)|\big]},\quad\tau=\pm1. 
\end{equation}
The corresponding Bogoliubov coefficients are given in Eqs.~\eqref{eq:coeff_bipartite_u}  and \eqref{eq:coeff_bipartite_v}. 
For IP polarization ($\sigma=x,y$), the $8\times8$ eigenvalue problem can 
in principle be solved analytically, but provides cumbersome expressions. For practical purposes we therefore
solve for the eigenproblem numerically. 

\begin{figure}[t!]
\begin{center}
\includegraphics[width=\columnwidth]{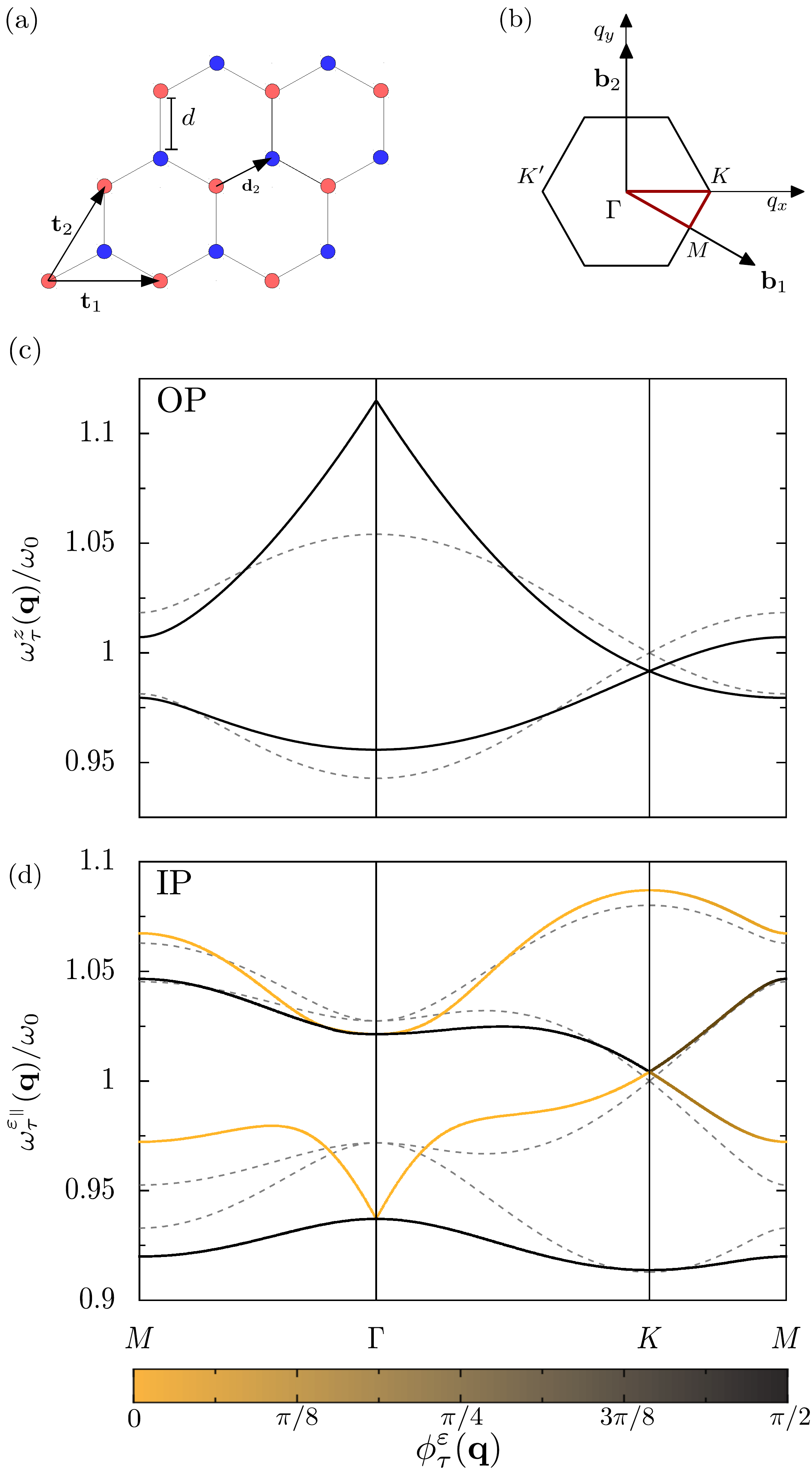}
\caption{\label{fig.Disp-honeycomb-z} (a) Sketch of a honeycomb lattice with primitive lattice vectors 
$\ta=d\, (\sqrt{3},0)$ and $\tb=d\, (\frac{\sqrt{3}}{2}, \frac 32)$, 
and basis vector $\mathbf{d}_{2}=d\, (\frac{\sqrt{3}}{2}, \frac 12)$. (b) Corresponding first Brillouin zone, with primitive reciprocal vectors 
$\mathbf{b}_1=\frac{2\pi}{3d}(\sqrt{3}, -1)$ and $\mathbf{b}_2=\frac{4\pi}{3d}(0, 1)$. 
(c) and (d) Quasistatic plasmonic band structure for (c) out-of-plane (OP) and (d) in-plane (IP) polarizations. 
The solid lines correspond to the full quasistatic dispersion, and the color code to the polarization angle \eqref{eq:polarization}. 
The dashed lines correspond to the nearest-neighbor approximation. 
Same parameters as in Fig.\ \ref{fig.Disp-square-z}.}
\end{center}
\end{figure}

To illustrate our method in the special case of bipartite lattices of near-field coupled metallic nanoparticles, we consider the celebrated honeycomb lattice sketched 
in Fig.~\ref{fig.Disp-honeycomb-z}(a). The corresponding 1BZ is shown in Fig.~\ref{fig.Disp-honeycomb-z}(b). 
Such a metasurface has been predicted \cite{Weick-PRL2013, Sturges-2DMat2015} to exhibit Dirac-like collective plasmonic modes at the $K$ and $K'$ points of the 1BZ, with appealing 
topological properties such as a nontrivial Berry phase (and its related topologically protected edge states \cite{Delplace2011-PRB}) or the absence of backscattering. Importantly, the results put forward in Refs.\ \cite{Weick-PRL2013, Sturges-2DMat2015} rely on short-range dipolar couplings between nearest neighbors alone. Moreover, Refs.~\cite{Weick-PRL2013, Sturges-2DMat2015} considered the case of orientated dipoles, relevant, e.g., for elongated metallic rods, while we consider here the case of spherical nanoparticles.  

The two plasmonic bands \eqref{eq:omega_z_bipartite} for $\sigma=z$ are plotted in Fig.~\ref{fig.Disp-honeycomb-z}(c) as solid lines. 
For comparison, we also show by dashed lines the plasmonic band structure with dipolar interactions between nearest neighbors only \cite{Weick-PRL2013}, given by Eq.~\eqref{eq:omega_z_bipartite} and replacing $f_{11}^{zz}(\q)$ and $f_{12}^{zz}(\q)$ by $f_{\mathrm{nn},11}^{zz}(\q)=0$ and $f_{\mathrm{nn},12}^{zz}(\q)=\sum_{j=1}^3\mathrm{e}^{\mathrm{i}\q\cdot \mathbf{e}_j}$, respectively.
Here, the vectors $\mathbf{e}_{1}=\mathbf{d}_{2}-\tb$, $\mathbf{e}_{2}=\mathbf{d}_{2}$, and $\mathbf{e}_{3}=\mathbf{d}_{2}-\ta$ connect 
a lattice site belonging to the red sublattice in Fig.\ \ref{fig.Disp-honeycomb-z}(a) to its three (blue) nearest neighbors. 
We observe in Fig.~\ref{fig.Disp-honeycomb-z}(c) the presence of a cusp for the upper ($\tau=+$) band when all quasistatic interactions are taken into account, while no cusp appears for the lower ($\tau=-$) band. 
Notice that the upper (lower) band corresponds to bright (dark) modes, where the two dipolar LSPs are in-phase (out-of-phase) within each unit cell. 

As can be seen from Fig.~\ref{fig.Disp-honeycomb-z}(c), the presence of a Dirac point at the $K$ point of the 1BZ located at $\mathbf{K}=\frac{4\pi}{3\sqrt{3}d}(1,0)$ is not ruled out by long-range interactions. Indeed, in the vicinity of the $K$ point, expanding the lattice sums to linear order in $|\mathbf{k}|$, where $\mathbf{q}=\mathbf{K}+\mathbf{k}$ with $|\mathbf{k}|\ll|\mathbf{K}|$, yields 
$f_{11}^{zz}(\mathbf{q})\simeq f_{11}^{zz}(\mathbf{K})\simeq -0.449$ and 
$f_{12}^{zz}(\mathbf{q})\simeq-1.16(k_x+\mathrm{i}k_y)d$. Therefore, in the 
weak-coupling regime $\Omega\ll\omega_0$, the dispersion \eqref{eq:omega_z_bipartite} is conical and reads
\begin{equation}
\label{eq:Dirac-cone}
\omega_\tau^{z}(\mathbf{k})\simeq\omega_0-\Omega|f_{11}^{zz}(\mathbf{K})|+\tau v^z|\mathbf{k}|, \quad\tau=\pm1,
\end{equation}
with the group velocity $v^z=1.16 \Omega d$. Comparing the dispersion above with the nearest-neighbor result \cite{Weick-PRL2013} $\omega_{\mathrm{nn},\tau}^{z}(\mathbf{k})=\omega_0+\tau v_\mathrm{nn}^z |\mathbf{k}|$ with $v_\mathrm{nn}^z=3\Omega d/2$, we see that the intrasublattice coupling $f_{11}^{zz}$ leads to an inconsequential redshift of the Dirac point frequency, while the intersublattice coupling $f_{12}^{zz}$ renormalizes the precise value of the group velocity. 

Since the Bogoliubov coefficients \eqref{eq:coeff_bipartite_v} are negligible as compared to the coefficients \eqref{eq:coeff_bipartite_u} close to the $K$ point, we can safely disregard the former, which amounts to performing the rotating wave approximation (RWA) \cite{SM}. Within this limit, 
the associated effective Hamiltonian reads in terms of the spinor operator $\hat\Psi_\mathbf{k}=(b_1^z(\mathbf{k}), b_2^z(\mathbf{k}))$ as 
$H_\mathrm{pl}^\mathrm{eff}=\sum_\mathbf{k}\hat \Psi_\mathbf{k}^\dagger \mathcal{H}^\mathrm{eff}_\mathbf{k}\hat\Psi_\mathbf{k}$, 
with the massless Dirac Hamiltonian
\begin{equation}
\label{eq:H_Dirac}
\mathcal{H}^\mathrm{eff}_\mathbf{k}=\left[\hbar\omega_0-\hbar\Omega|f_{11}^{zz}(\mathbf{K})|\right]\mathbb{1}_2-\hbar v^z \boldsymbol{\sigma}\cdot\mathbf{k},
\end{equation}
where $\mathbb{1}_n$ is the $n\times n$ identity matrix and $\boldsymbol{\sigma}=(\sigma_x, \sigma_y, \sigma_z)$ is the vector of Pauli matrices
\begin{equation}
\sigma_x=
\begin{pmatrix}
0 & 1 \\
1 & 0
\end{pmatrix}\quad
\sigma_y=
\begin{pmatrix}
0 & -\mathrm{i} \\
\mathrm{i} & 0
\end{pmatrix}\quad
\sigma_z=
\begin{pmatrix}
1 & 0 \\
0 & -1
\end{pmatrix}
\end{equation}
acting on the sublattice pseudospin $1/2$. The corresponding plasmonic eigenstates [with eigenfrequencies given by Eq.~\eqref{eq:Dirac-cone}] read
\begin{equation}
\label{eq:spinor}
|\psi_\tau^z(\mathbf{k})\rangle=\frac{1}{\sqrt{2}}
\begin{pmatrix}
1 \\
-\tau\mathrm{e}^{\mathrm{i}\theta(\mathbf{k})}
\end{pmatrix},\quad\tau=\pm1,
\end{equation} 
where $\theta(\mathbf{k})$ is the angle between the wave vector $\mathbf{k}$ and the $x$ axis, such that $\tan{\theta(\mathbf{k})}=k_y/k_x$.  
The spinors \eqref{eq:spinor} are also eigenstates of the chirality (or helicity) operator~\cite{Allain2011}, $\mathcal{C}=\boldsymbol{\sigma}\cdot\hat k$, which corresponds to the projection of the pseudospin operator onto the wave vector direction, with eigenvalues $C=-\tau$. Since $[\mathcal{H}^\mathrm{eff}_\mathbf{k}, \mathcal{C}]=0$, chirality is conserved. 
As a consequence, collective plasmonic modes close to the $K$ point show similar effects as electrons in graphene \cite{Weick-PRL2013}, such as the absence of backscattering off smooth inhomogeneities (since 
$\langle \psi_\tau^z(-\mathbf{k})|\psi_\tau^z(\mathbf{k})\rangle=0$~\cite{Allain2011}), the related Klein tunneling phenomenon, or a Berry flux of $-\pi$. 	
Note that the spinor eigenstates \eqref{eq:spinor} do not depend on the details of the Hamiltonian \eqref{eq:H_Dirac}, such as, e.g., the precise value of the group velocity $v^z$, but only on its matrix structure. Henceforth, the properties mentioned above do not depend on the level of approximation one may use (nearest-neighbor coupling vs long-range quasistatic interactions). 
In addition, we will demonstrate in Sec.~\ref{sec:rad_frequency_shifts} that the retarded part of the dipolar coupling 
does not modify the above physics. 
Therefore, the long-range character of the dipole--dipole interaction does not rule out the massless Dirac nature of the plasmonic quasiparticles in the vicinity of the $K$ point, and the nearest-neighbor approximation is sufficient in catching the relevant physics. The same conclusion applies to the inequivalent Dirac point located at $K'$. 

We now turn to the description of the IP polarized plasmonic modes. We show in Fig.\ \ref{fig.Disp-honeycomb-z}(d) the plasmonic band structure obtained numerically for $\sigma=x,y$ as solid lines. 
The color code corresponding to the polarization angle \eqref{eq:polarization} reveals that two bands correspond to purely transverse plasmons, and two other bands to purely longitudinal plasmons along the high-symmetry axes $\Gamma K$ and $\Gamma M$. 
We also plot
for comparison the dispersion relations with nearest-neighbor coupling only. In the latter case, the intrasublattice sums $f^{\sigma\sigma'}_{\mathrm{nn},11}(\q)$  vanish, so that the $8\times 8$ system 
given by Eq.~\eqref{Hopfield-coeff} simplifies and can be straightforwardly solved analytically. The resulting four dispersion relations read 
\begin{equation}
\label{eq:omega_xy_bipartite}
\omega_{\mathrm{nn},\tau}^{\varepsilon_\parallel}(\q)=
\begin{cases}
\displaystyle
\omega_{0}\left[1+\sqrt{2}\frac{\Omega}{\omega_{0}}\sqrt{\mathcal{G}_1(\q)\pm\sqrt{\mathcal{G}_2(\q)}}\right]^{1/2},
\vspace{.1truecm}
\\
\displaystyle
\omega_{0}\left[1-\sqrt{2}\dfrac{\Omega}{\omega_{0}}\sqrt{\mathcal{G}_1(\q)\pm\sqrt{\mathcal{G}_2(\q)}}\right]^{1/2}, 
\end{cases}
\end{equation}
with
\begin{equation}
\mathcal{G}_1(\q)=|f^{xx}_{\mathrm{nn},12}(\q)|^2+|f^{yy}_{\mathrm{nn},12}(\q)|^2+2|f^{xy}_{\mathrm{nn},12}(\q)|^2
\end{equation}
and
\begin{align}
\mathcal{G}_2(\q)=&\;\left[|f^{xx}_{\mathrm{nn},12}(\q)|^2-|f^{yy}_{\mathrm{nn},12}(\q)|^2\right]^2\nonumber\\
&+4|f^{xy}_{\mathrm{nn},12}(\q)|^2\left[|f^{xx}_{\mathrm{nn},12}(\q)|^2+|f^{yy}_{\mathrm{nn},12}(\q)|^2\right]\nonumber\\
&+8\, \mathrm{Re}\left\{f^{xx}_{\mathrm{nn},12}(\q)f^{yy}_{\mathrm{nn},12}(\q){\left[f^{xy}_{\mathrm{nn},12}(\q)\right]^*}^2\right\}.
\end{align}
In the expressions above, the intersublattice sums within the nearest-neighbor approximation read 
$f_{\mathrm{nn},12}^{xx}(\q)=\mathrm{e}^{\mathrm{i}\q\cdot\mathbf{e}_{1}}
-\frac{5}{4}(\mathrm{e}^{\mathrm{i}\q\cdot\mathbf{e}_{2}}
+\mathrm{e}^{\mathrm{i}\q\cdot\mathbf{e}_{3}})$, 
$f_{\mathrm{nn},12}^{yy}(\q)=-2\mathrm{e}^{\mathrm{i}\q\cdot\mathbf{e}_{1}}
+\frac{1}{4}(\mathrm{e}^{\mathrm{i}\q\cdot\mathbf{e}_{2}}
+\mathrm{e}^{\mathrm{i}\q\cdot\mathbf{e}_{3}})$,
and
$f_{\mathrm{nn},12}^{xy}(\q)=
\frac{3\sqrt{3}}{4}(\mathrm{e}^{\mathrm{i}\q\cdot\mathbf{e}_{3}}
-\mathrm{e}^{\mathrm{i}\q\cdot\mathbf{e}_{2}})$.
As can be seen from Fig.~\ref{fig.Disp-honeycomb-z}(d), the nearest-neighbor approximation is sufficient to qualitatively describe the plasmonic dispersion relations, apart from the second less energetic band for wave numbers close to the $\Gamma$ point, where a cusp appears in the full 
quasistatic band structure. 
Importantly, we note the presence of two inequivalent conical intersections (where the band degeneracy point occurs at the frequency $\omega_0$) at the $K$ and $K'$ points of the 1BZ for IP polarized modes. 
In the vicinity of the $K$ point, we find for the second and third bands $\omega_{\mathrm{nn},\tau}^{\varepsilon_\parallel}(\mathbf{k})\simeq\omega_0\pm v_\mathrm{nn}^{\varepsilon_\parallel}|\mathbf{k}|$, with $v_\mathrm{nn}^{\varepsilon_\parallel}=9\Omega d/4$. 
The presence of conical intersections for IP polarized modes has been reported by Han \textit{et al.}\ \cite{Han-PRL2009} using a numerical solution to Maxwell's equations. Our transparent method allows us to analytically describe such a complex band structure hosting Dirac-like bosonic modes.

\subsubsection{Tripartite lattices}
\label{sec:tripartite}

For tripartite lattices ($\mathcal{S}=3$), the $6\times 6$ system obtained from Eq.\ \eqref{Hopfield-coeff} for $\sigma=z$ can still be solved analytically, yielding the three plasmonic bands 
\begin{subequations}
\label{eq:omega_z_tripartite}
\begin{align}
\omega_\tau^z(\q)=
\omega_0\bigg\{
1+2\frac{\Omega}{\omega_0}\Big[f^{zz}_{11}(\q)+s_+(\q)+s_-(\q)\Big]
\bigg\}^{1/2},
\end{align}
\begin{align}
\omega_\tau^z(\q)=&\;
\omega_0\bigg\{
1+\frac{\Omega}{\omega_0}\Big[2f^{zz}_{11}(\q)-s_+(\q)-s_-(\q)
\nonumber\\
&\pm\mathrm{i}\, \sqrt{3}\big(s_+(\q)-s_-(\q)\big)\Big]
\bigg\}^{1/2},
\end{align}
\end{subequations}
where
\begin{align}
s_\pm(\q)=
\left\{
\Pi^{zz}(\q)
\pm\mathrm{i}\,
\sqrt{\left[\frac{\Sigma^{zz}(\q)}{3}\right]^3
-\big[
\Pi^{zz}(\q)
\big]^2
}
\right\}^{1/3}
\end{align}
with 
\begin{equation}
\label{eq:Sigmazz}
\Sigma^{zz}(\q)=|f_{12}^{zz}(\q)|^2+|f_{13}^{zz}(\q)|^2+|f_{23}^{zz}(\q)|^2
\end{equation}
 and 
 \begin{equation}
 \label{eq:Pizz}
 \Pi^{zz}(\q)=\mathrm{Re}\left\{f_{12}^{zz}(\q){f_{13}^{zz}}^*(\q)f_{23}^{zz}(\q)\right\}.
 \end{equation}
For $\sigma=x,y$, the $12\times 12$ eigensystem is solved numerically.

\begin{figure}[t!]
\begin{center}
\includegraphics[width=\columnwidth]{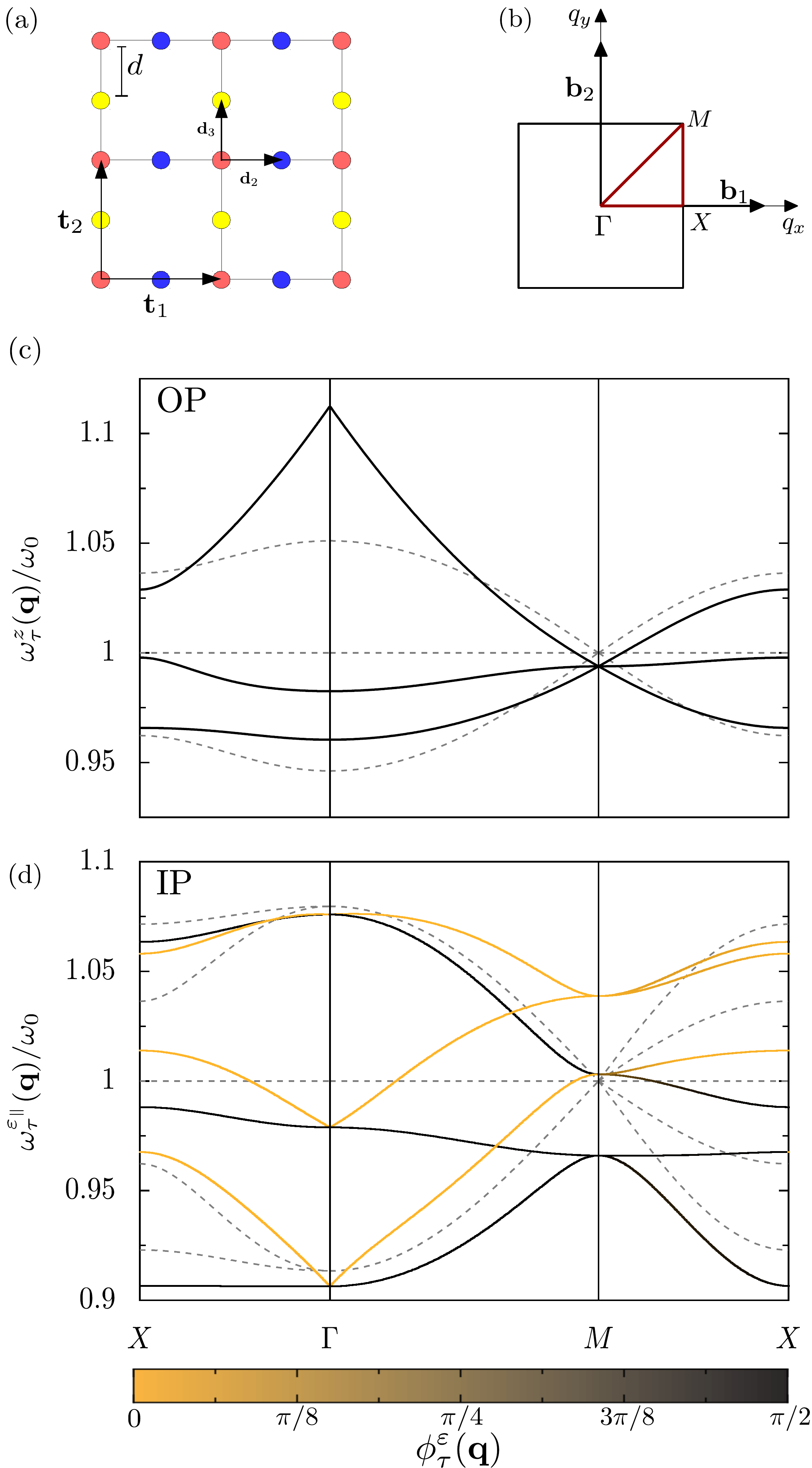}
\caption{\label{fig.Disp-lieb-z} (a) Sketch of a Lieb lattice with primitive lattice vectors 
$\ta=d\, (2,0)$ and $\tb=d\, (0, 2)$, 
and basis vectors $\mathbf{d}_{2}=\ta/2$ and $\mathbf{d}_{3}=\tb/2$. (b) Corresponding first Brillouin zone, with primitive reciprocal vectors 
$\mathbf{b}_1=\frac{\pi}{d}(1, 0)$ and $\mathbf{b}_2=\frac{\pi}{d}(0, 1)$. 
(c) and (d) Quasistatic plasmonic band structure for (c) out-of-plane (OP) and (d) in-plane (IP) polarizations. 
The dashed and solid lines correspond, respectively, to nearest-neighbor and long-range couplings. 
Color code of the solid lines: polarization angle \eqref{eq:polarization}. 
Same parameters as in Fig.~\ref{fig.Disp-square-z}.}
\end{center}
\end{figure}

As an illustration, we consider in the remainder of this section the Lieb lattice sketched in Fig.\ \ref{fig.Disp-lieb-z}(a). Its corresponding 1BZ is shown in Fig.\ \ref{fig.Disp-lieb-z}(b). Such a lattice, together with the kagome one \cite{SM}, is known to display a flat (i.e., nondispersive) band together with conical dispersions in tight-binding models with nearest-neighbor interactions
(see, e.g., Refs.~\cite{Bergman-PRB2008, Guo-PRB2009, Weeks-PRB2010, Nita2013_PRB, Palumbo-PRB2015}). 
It is therefore of interest to study if these features survive in the case of plasmonic metasurfaces, where the long-range nature of the dipolar interaction might qualitatively change the band structure. 

We plot in Fig.\ \ref{fig.Disp-lieb-z}(c) the three dispersion relations \eqref{eq:omega_z_tripartite} for the Lieb lattice and for OP polarization as solid lines. For comparison, we also show by dashed lines the plasmonic band structure within the nearest-neighbor approximation, 
for which we have $f_{\mathrm{nn}, 11}^{zz}(\q)=f_{\mathrm{nn}, 22}^{zz}(\q)=f_{\mathrm{nn}, 33}^{zz}(\q)=f_{\mathrm{nn}, 23}^{zz}(\q)=0$, $f_{\mathrm{nn}, 12}^{zz}(\q)=2\cos{(q_xd)}$, and $f_{\mathrm{nn}, 13}^{zz}(\q)=2\cos{(q_yd)}$. 
Within this approximation, the resulting band structure is 
given by $\omega_{\mathrm{nn},\tau}^{z}(\q)=\omega_0[1+\tau\frac{4\Omega}{\omega_0}\sqrt{\cos^2{(q_xd)}+\cos^2{(q_yd)}}]^{1/2}$ and is
characterized by the presence of a flat band at $\omega_0$ ($\tau=0$) and two dispersive bands ($\tau=\pm1$). 
For wave vectors $\mathbf{k}$ in the vicinity of the $M$ point of the 1BZ located at $\mathbf{M}=\frac{\pi}{2d}(1, 1)$, these latter form a conical dispersion 
$\omega_{\mathrm{nn},\pm 1}^{z}(\kp)\simeq\omega_{0}\pm v_\mathrm{nn}^z |\kp|$, with group velocity $v^z_\mathrm{nn}=2\Omega d$.

As shown in Fig.\ \ref{fig.Disp-lieb-z}(c), the long-range dipolar interactions affect differently the topological features of the plasmonic band structure 
described above (compare solid and dashed lines). Indeed, while the presence of a conical dispersion in the vicinity of the $M$ point is not ruled out by long-range interactions (as is the case for the honeycomb array, see Sec.~\ref{sec:bipartite}), the band which is flat in the whole 1BZ within the nearest-neighbor approximation becomes dispersive, and is only locally flat close to the $M$ point. One can understand these features by expanding the band structure \eqref{eq:omega_z_tripartite} in the vicinity of the $M$ point. With $\mathbf{q}=\mathbf{M}+\mathbf{k}$, to linear order in $|\mathbf{k}|\ll|\mathbf{M}|$ we find 
$f_{ss}^{zz}(\mathbf{q})\simeq f_{ss}^{zz}(\mathbf{M})\simeq-0.331$, 
$f_{12}^{zz}(\mathbf{q})\simeq-1.65 k_xd$, 
$f_{13}^{zz}(\mathbf{q})\simeq-1.65 k_yd$, and 
$f_{23}^{zz}(\mathbf{q})\simeq0$, resulting in 
\begin{equation}
\label{eq:Lieb_eff_disp}
\omega_{\tau}^{z}(\kp)\simeq\omega_{0}-\Omega|f_{11}(\mathbf{M})|+\tau v^z |\kp|, \quad\tau=0, \pm1, 
\end{equation}
i.e., a band $\tau=0$ which is locally flat, 
and two conical dispersions $\tau=\pm1$ with renormalized group velocity $v^z=1.65 \Omega d$. 

Interestingly, the plasmonic Hamiltonian within the RWA for modes polarized in the $z$ direction takes the form $H_\mathrm{pl}^\mathrm{eff}=\sum_\mathbf{k}\hat \Psi_\mathbf{k}^\dagger \mathcal{H}^\mathrm{eff}_\mathbf{k}\hat\Psi_\mathbf{k}$, with 
$\hat\Psi_\mathbf{k}=(b_1^z(\mathbf{k}), b_2^z(\mathbf{k}), b_3^z(\mathbf{k}))$ and where
\begin{equation}
\label{eq:H_eff_Lieb}
\mathcal{H}^\mathrm{eff}_\mathbf{k}=\left[\hbar\omega_0-\hbar\Omega|f_{11}^{zz}(\mathbf{M})|\right]\mathbb{1}_3-\hbar v^z \mathbf{S}\cdot\mathbf{k},
\end{equation}
with $\mathbf{S}=(S_x, S_y, S_z)$.
Here, the pseudospin-$1$ matrices (corresponding to the three sublattices of the Lieb lattice) are defined as 
\begin{equation}
S_x=
\begin{pmatrix}
0 & 1 & 0 \\
1 & 0 & 0 \\
0 & 0 & 0
\end{pmatrix}\quad
S_y=
\begin{pmatrix}
0 & 0 & 1 \\
0 & 0 & 0 \\
1 & 0 & 0
\end{pmatrix}\quad
S_z=
\begin{pmatrix}
0 & 0 & 0 \\
0 & 0 & -\mathrm{i} \\
0 & \mathrm{i} & 0
\end{pmatrix}
\end{equation}
and fulfill the angular momentum algebra $[S_i, S_j]=\mathrm{i}\epsilon_{ijk}S_k$, with $\epsilon_{ijk}$ the Levi-Civita symbol. The matrices $S_x$, $S_y$, and $S_z$ (which correspond, respectively, to the Gell-Mann matrices $\lambda_1$, $\lambda_4$, and $\lambda_7$ \cite{Gell-Mann-PR1962, Arfken}) therefore correspond to a three-dimensional representation of the special unitary group $SU(2)$. However, contrary to the Pauli matrices, they do not form a Clifford algebra (i.e., $\{S_i, S_j\}\neq 2\delta_{ij}\mathbb{1}_3$), so that Eq.~\eqref{eq:H_eff_Lieb} does not correspond to a massless Dirac Hamiltonian~\cite{Goldman-PRA2011}, despite presenting a conical spectrum.
The eigenspinors corresponding to the Hamiltonian \eqref{eq:H_eff_Lieb} with eigenfrequencies \eqref{eq:Lieb_eff_disp}, and characterized by a vanishing Berry phase, read
\begin{equation}
|\psi_\tau^z(\kp)\rangle=\frac{1}{\sqrt{2}}
\begin{pmatrix}
\tau^2  \\
-\tau\cos{\theta(\kp)}+\sqrt{2}(1-\tau^2)\sin{\theta(\kp)}  \\
-\tau\sin{\theta(\kp)}-\sqrt{2}(1-\tau^2)\cos{\theta(\kp)}
\end{pmatrix} 
\end{equation}
for $\tau=0, \pm1$ and are nevertheless eigenstates of the helicity operator 
$\mathbf{S}\cdot\hat k$ with eigenvalues $-\tau$, 
 so that backscattering is suppressed for the two dispersive bands with $\tau=\pm1$ ($\langle \psi_{\pm1}^z(-\mathbf{k})|\psi_{\pm1}^z(\mathbf{k})\rangle=0$). The conclusion above is not modified when we take into account the retarded part of the dipolar interaction (see Sec.~\ref{sec:rad_frequency_shifts}).

We show in Fig.\ \ref{fig.Disp-lieb-z}(d) the plasmonic dispersion relations calculated numerically for $\sigma=x,y$ as solid lines. We further plot, for comparison, the numerical results with nearest-neighbor couplings only. We observe from the figure that for IP polarization, the long-range dipolar interactions completely reconstruct the plasmonic band structure. Notably, the topological features (flat bands and conical dispersions) occurring in the nearest-neighbors coupling approximation are not preserved when long-range interactions are included.

\section{Perturbative treatment of the retardation effects onto the plasmonic band structure}
\label{sec:rad_frequency_shifts}
We now consider the effects of the photonic environment alone [encapsulated in the Hamiltonian \eqref{H-ph}] onto the collective plasmonic excitations supported by our generic metasurface of ordered metallic nanoparticles. The plasmon--photon interaction [cf.\ Eq.\ \eqref{H-plph}] leads to two effects crucially affecting the quasistatic band structure discussed above: (i) the photon-induced frequency shifts which we unveil in the present section and resulting from retardation effects in the dipole--dipole interaction renormalize the quasistatic plasmonic dispersion; (ii) moreover, the spontaneous decay of plasmons into free photons leads to a finite radiative lifetime of the collective excitations, and consequently to a broadening of the plasmon lines (see Sec.\ \ref{sec:rad_linewidths}).

We start our analysis of the effects of the photonic environment onto the quasistatic plasmonic dispersion by considering the radiative frequency shifts induced by the light--matter coupling. Along the lines of Refs.~\cite{downi17a_preprint, downi18_JPCM, downi18_EPJB}, we 
treat the plasmon--photon coupling Hamiltonian \eqref{H-plph} to second order in standard nondegenerate perturbation theory, yielding for the collective plasmon energy levels the result 
$E_{n_\tau^\varepsilon(\q)}=n_\tau^\varepsilon(\q)\hbar\omega_\tau^\varepsilon(\q)+E_{n_\tau^\varepsilon(\q)}^{(1)}+E_{n_\tau^\varepsilon(\q)}^{(2)}$.
The first term in the right-hand side of the equation above corresponds to the energy levels of the unperturbed Hamiltonian $H_\tau^\varepsilon(\q)$ [cf.\ Eq.\ \eqref{Plasmonic-H-diagonal}]. The first-order (in $H_\mathrm{pl\textrm{-}ph}$)
correction stems from the diamagnetic term in Eq.\ \eqref{H-plph} (proportional to the vector potential squared) and reads
$E_{n_\tau^\varepsilon(\q)}^{(1)}=2\pi\mathcal{SN}\hbar\omega_0^2a^3\mathcal{V}^{-1}\sum_{\mathbf{k}}\omega_\mathbf{k}^{-1}$.
Since the latter expression does not depend on the quantum number $n_\tau^\varepsilon(\q)$, it does not participate in the renormalization of the collective plasmon frequency, and merely represents an irrelevant global energy shift. The second-order correction to $n_\tau^\varepsilon(\q)\hbar\omega_\tau^\varepsilon(\q)$ arising from the first (paramagnetic) term on the right-hand side of Eq.~\eqref{H-plph} reads
\begin{align}
\label{eq:E_rad_2}
E_{n_\tau^\varepsilon(\q)}^{(2)}=&\;\pi\hbar\omega_0^3\frac{a^3}{\mathcal{V}}
\sum_{\mathbf{k},\hat\lambda_\mathbf{k}}\frac{1}{\omega_\mathbf{k}}
\Bigg\{
\frac{n_\tau^\varepsilon(\q)}{\omega_\tau^\varepsilon(\q)-\omega_\mathbf{k}}
\nonumber\\
&\times \left|F_{\mathbf{k},\q}^-\sum_{s\sigma}(\hat\sigma\cdot\hat\lambda_\mathbf{k})\,
\mathrm{e}^{-\mathrm{i}(\q-\mathbf{k})\cdot\mathbf{d}_s}\,P_{\tau s}^{\varepsilon\sigma}(\q)\right|^2
\nonumber\\
&-\frac{n_\tau^\varepsilon(\q)+1}{\omega_\tau^\varepsilon(\q)+\omega_\mathbf{k}}
\nonumber\\
&\times
\left|F_{\mathbf{k},\q}^+\sum_{s\sigma}(\hat\sigma\cdot\hat\lambda_\mathbf{k})\,
\mathrm{e}^{-\mathrm{i}(\q+\mathbf{k})\cdot\mathbf{d}_s}\,P_{\tau s}^{\varepsilon\sigma}(\q)\right|^2
\Bigg\},
\end{align}
where the summation over $\mathbf{k}$ excludes the singular term for which $\omega_\mathbf{k}=\omega_\tau^\varepsilon(\q)$.
In Eq.\ \eqref{eq:E_rad_2}, we have defined the array factor
\begin{equation}
\label{eq:array_factor}
F_{\mathbf{k},\q}^\pm=\frac{1}{\sqrt{\mathcal{N}}}\sum_{\mathbf{R}}\mathrm{e}^{\mathrm{i}(\q\pm\mathbf{k})\cdot\mathbf{R}}
\end{equation}
and 
\begin{equation}
\label{eq:P}
P_{\tau s}^{\varepsilon\sigma}(\q)=u_{\tau s}^{\varepsilon\sigma}(\q)+v_{\tau s}^{\varepsilon\sigma}(\q).
\end{equation}

In the large-metasurface limit (number of unit cells $\mathcal{N}\gg1$), the modulus squared of the array factor above entering Eq.~\eqref{eq:E_rad_2} takes the simpler form
\begin{equation}
\label{eq:array_factor_infinite}
|F_{\mathbf{k},\q}^\pm|^2\simeq
\frac{(2\pi)^2}{|\mathbf{t}_1\times\mathbf{t}_2|}\delta(q_x\pm k_x)\delta(q_y\pm k_y), 
\end{equation}
so that the radiative frequency shift, defined through the relation 
$\delta_\tau^\varepsilon(\q)=[E_{n_\tau^\varepsilon(\q)+1}-E_{n_\tau^\varepsilon(\q)}]/\hbar-\omega_\tau^\varepsilon(\q)$, 
is given by
\begin{align}
\label{eq:rad_shift_general}
\delta_\tau^\varepsilon(\q)=&\;\pi\omega_0^3\frac{a^3}{\mathcal{V}}\sum_{\mathbf{k},\hat\lambda_\mathbf{k}}
\frac{\left|\sum_{s\sigma}(\hat\sigma\cdot\hat\lambda_\mathbf{k})P_{\tau s}^{\varepsilon\sigma}(\q)\right|^2}{\omega_\mathbf{k}}
\nonumber\\
&\times\left[
\frac{|F_{\mathbf{k}, \mathbf{q}}^-|^2}{\omega_\tau^\varepsilon(\q)-\omega_\mathbf{k}}
-
\frac{|F_{\mathbf{k}, \mathbf{q}}^+|^2}{\omega_\tau^\varepsilon(\q)+\omega_\mathbf{k}}
\right]
\end{align}
and scales with the nanoparticle sizes as $a^3$, to leading order in the coupling constant $\Omega/\omega_0$ [cf.\ Eq.~\eqref{eq:Omega}].

The analytical result above is valid for both IP and OP plasmon polarizations, and depends on the quasistatic plasmon dispersion $\omega_\tau^\varepsilon(\q)$ and the Bogoliubov coefficients [cf.\ Eq.~\eqref{eq:P}]. Equation \eqref{eq:rad_shift_general} can be expicitly evaluated~\cite{SM}, and we find
\begin{align}
\label{eq:delta_z}
\delta_\tau^z(\q)=&\;\frac{\pi\omega_0^3a^3|\q|}{|\mathbf{t}_1\times\mathbf{t}_2|[\omega_\tau^z(\q)]^2}
\left|\sum_{s}P_{\tau s}^{zz}(\q)\right|^2
\nonumber\\
&\times
\left[
1-\frac{c|\q|}{\sqrt{\left(c|\q|\right)^2-\left[\omega_\tau^z(\q)\right]^2}}\Theta\big(c|\q|-\omega_\tau^z(\q)\big)
\right].
\end{align}
and 
\begin{align}
\label{eq:delta_inplane}
\delta_\tau^{\varepsilon_\parallel}(\q)=&\,-\frac{\pi\omega_0^3a^3|\q|}{|\mathbf{t}_1\times\mathbf{t}_2|
[\omega_\tau^{\varepsilon_\parallel}(\q)]^2}
\nonumber\\
&\times
\left\{
\left|\sum_{s}\left[\frac{q_x}{|\q|}P_{\tau s}^{\varepsilon_\parallel x}(\q)
+\frac{q_y}{|\q|}P_{\tau s}^{\varepsilon_\parallel y}(\q)\right]\right|^2
\right.
\nonumber\\
&\times
\left[
1-\frac{c|\q|}{\sqrt{\left(c|\q|\right)^2-\left[\omega_\tau^{\varepsilon_\parallel}(\q)\right]^2}}\Theta\big(c|\q|-\omega_\tau^{\varepsilon_\parallel}(\q)\big)
\right]
\nonumber\\
&+
\left[
\left|\sum_{s}P_{\tau s}^{\varepsilon_\parallel x}(\q)\right|^2
+
\left|\sum_{s}P_{\tau s}^{\varepsilon_\parallel y}(\q)\right|^2
\right]
\nonumber\\
&\times
\left.
\frac{\left[\omega_\tau^{\varepsilon_\parallel}(\q)\right]^2}{c|\q|\sqrt{\left(c|\q|\right)^2-\left[\omega_\tau^{\varepsilon_\parallel}(\q)\right]^2}}\Theta\big(c|\q|-\omega_\tau^{\varepsilon_\parallel}(\q)\big)
\right\}
.
\end{align}
for the radiative shifts of the OP and IP plasmonic modes, respectively.
Notably, the calculation leading to the above results does not need the introduction of an ultraviolet frequency cutoff for the photonic degrees of freedom [which prevent to take into account photons with a wavelength smaller than the nanoparticle size, cf.\ the dipolar approximation in Eq.\ \eqref{H-plph}], as is the case for nanoparticle dimers \cite{downi17a_preprint} and linear chains~\cite{downi18_JPCM, downi18_EPJB}. We have checked that the inclusion of such a cutoff, of the order of $k_\mathrm{c}\simeq1/a$, does not significantly modify the results \eqref{eq:delta_z} 
and \eqref{eq:delta_inplane}. 
Notice also that Eqs.~\eqref{eq:delta_z} and \eqref{eq:delta_inplane} are not periodic in reciprocal space since we consider only the interaction between the collective plasmons and photons  for which the associated light cone belongs to the 1BZ. Finally, we note from Eqs.~\eqref{eq:delta_z} and \eqref{eq:delta_inplane} that $\delta_\tau^\varepsilon(\mathbf{0})=0$, so that the dispersion relation at the $\Gamma$ point is not renormalized by the light--matter interaction. 

\begin{figure*}[t]
\begin{center}
\includegraphics[width=\textwidth]{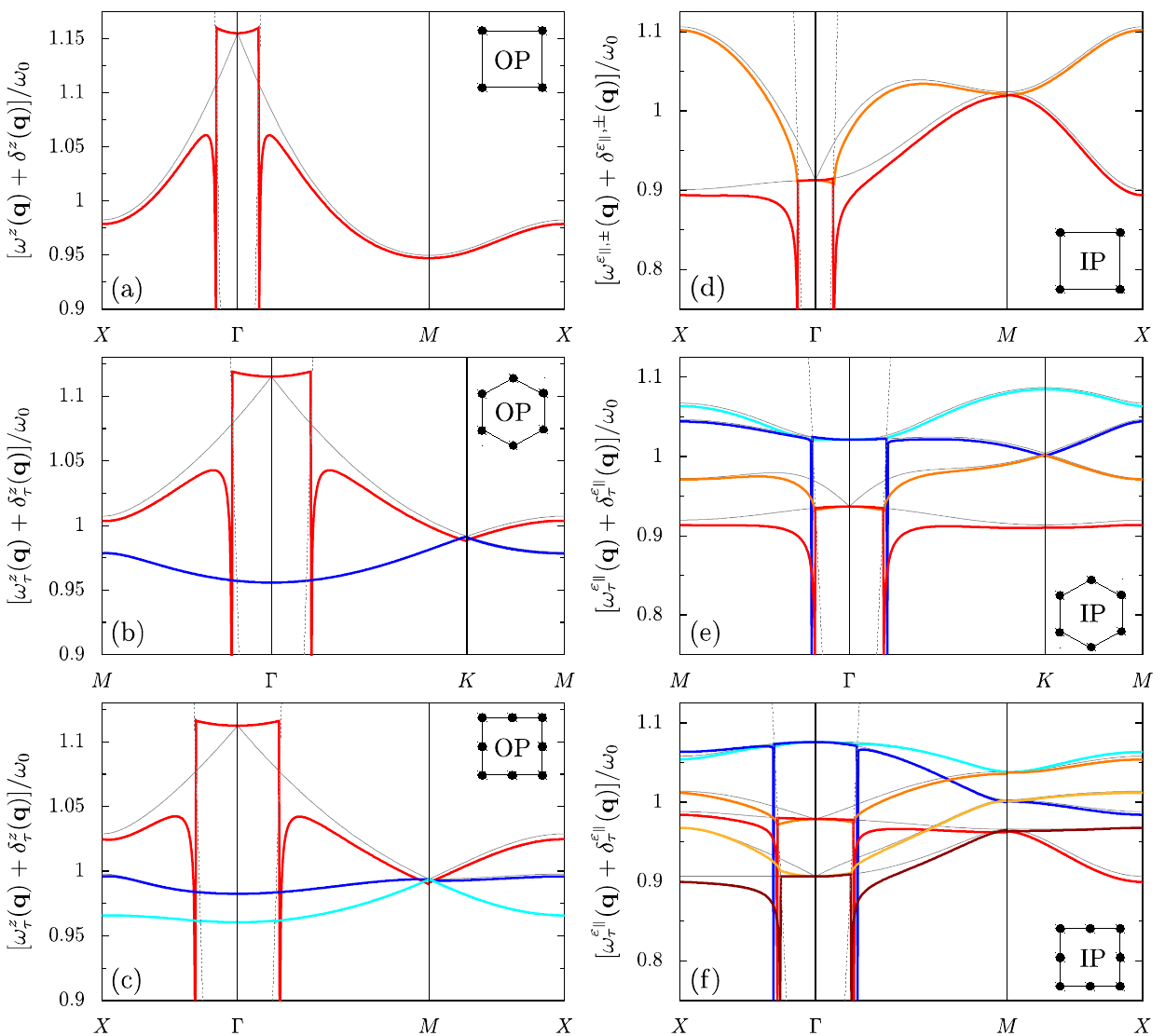}
\caption{\label{fig.Disp_retarded} The colored lines display the plasmonic band structure $\omegatau+\delta_{\tau}^{\varepsilon}(\q)$ (scaled by $\omega_{0}$) including retardation effects and the corresponding radiative frequency shift \eqref{eq:rad_shift_general} for (a) and (d) the square, (b) and (e) the honeycomb, and (c) and (f) the Lieb lattices. (a)--(c) [(d)--(f)] show the OP [IP] modes. For comparison purposes, the thin gray lines reproduce the corresponding quasistatic dispersion relations $\omegatau$, while the dashed gray lines depict the light cone within each 1BZ. In the figure, $d=3a$ and $k_{0}a=0.15$.}
\end{center}
\end{figure*}

We plot in Figs.\ \ref{fig.Disp_retarded}(a)--\ref{fig.Disp_retarded}(c) the plasmonic band structure for OP-polarized modes $\omega_\tau^z(\q)+\delta_{\tau}^{z}(\q)$ including Eq.~\eqref{eq:delta_z} for (a) the square, (b) the honeycomb, and (c) the Lieb lattices along the high symmetry lines of their respective 1BZ. 
As shown in Fig.~\ref{fig.Disp_retarded}(a) for the square lattice, the radiative frequency shift $\delta_{\tau}^{z}(\q)$ induces an important renormalization of the quasistatic band structure, which is of purely transverse nature [see Fig.~\ref{fig.Disp-square-z}(c)]. 
A similar effect is also observed for other simple Bravais lattices, such as the rectangular and hexagonal ones \cite{SM}.
Comparing the retarded dispersion relation (red line) to the quasistatic one (thin gray line), we first observe that the retarded one
diverges at the intersection of the quasistatic band structure with the light cone (dashed lines). 
A polaritonic singularity arising from the strong light--matter interaction was already reported in the literature for the transverse modes in 1D plasmonic chains by means of involved numerical calculations based on the fully retarded solutions to Maxwell's equations~\cite{Weber-PRB2004, Citrin-Nanolett2004, Simovski-PRE2005, Citrin-OptLett2006, Koenderink-PRB2006, Markel-PRB2007, Fung-OptLett2007, Petrov-PRA2015}. Recently, studies~\cite{downi18_JPCM, downi18_EPJB} using the open quantum system approach which we use in this work have shown a similar divergence occurring in the dispersion relations of transverse plasmonic modes in chains. Importantly, the results presented in Ref.\ \cite{downi18_JPCM} show a good qualititative agreement with numerical electromagnetic calculations for regular nanoparticle chains. We here point out that the singularities observed in Fig.~\ref{fig.Disp_retarded} arise from calculations based on a second-order perturbative treatment of the light--matter coupling. Consequently, important variations from the LSP resonance frequency $\omega_{0}$ should be treated carefully. Notice also that the renormalized dispersion relation diverges as the inverse of a square root [see Eq.~\eqref{eq:delta_z}] instead of logarithmically as is the case for 1D arrays~\cite{downi18_JPCM, downi18_EPJB}.

As can be seen from Fig.\ \ref{fig.Disp_retarded}(a), taking into account the radiative shift implies that the cusp appearing at the $\Gamma$ point of the 1BZ within the quasistatic approximation disappears (compare the red and gray lines in the figure). In the vicinity of the $\Gamma$ point ($|\q|d\ll 1$), Eq.~\eqref{eq:delta_z} applied to the square lattice reduces (considering $\Omega/\omega_{0}\ll 1)$ to $\delta^{z}(\q)\simeq 2\pi\Omega|\q|d+\mathcal{O}(|\q| d)^3$. Such a linear $|\q|$-dependence cancels out exactly the one of $\omega^{z}(\q)$ in this regime of parameters [see Eq.~\eqref{eq:Bethe-Peierls_square}]. We hence find 
$\omega^{z}(\q)+\delta^{z}(\q)\simeq \omega_{0}+\Omega[4+\sqrt{2}\pi+(|\q|d)^{2}(\pi/\sqrt{2}-1)]$
close to the $\Gamma$ point, leading to a quadratic dependence of the dispersion relation. The results presented in Fig.~\ref{fig.Disp_retarded}(a) thus demonstrate that considering retardation effects is crucial for the study of the collective plasmonic modes in metasurfaces of near-field coupled nanoparticles, since the dispersion relations are qualitatively affected by the interactions with free photons. 
Such renormalization effects are not that prominent in 1D plasmonic systems, where, apart from the pronounced renormalization of the dispersion relation at the crossing of the light cone, the band structure is qualitatively unaffected by retardation effects 
\cite{downi18_JPCM, downi18_EPJB, Weber-PRB2004, Citrin-Nanolett2004, Simovski-PRE2005, Citrin-OptLett2006, Koenderink-PRB2006, Markel-PRB2007, Fung-OptLett2007, Petrov-PRA2015}.

We show in Fig.\ \ref{fig.Disp_retarded}(b) the band structure including retardation effects (colored lines) of the OP plasmonic modes for the honeycomb lattice. For comparison, we also reproduce as gray solid 
lines the transverse-polarized quasistatic band structure shown in Fig.~\ref{fig.Disp-honeycomb-z}(c). 
The upper band ($\tau=+1$) shows a similar profile as that in Fig.~\ref{fig.Disp_retarded}(a). It displays a divergence at the intersection 
between the light cone (dashed lines) and the quasistatic dispersion relation. In addition, it does not present a cusp at the $\Gamma$ point. In contrast, the lower band ($\tau=-1$) does not experience a noticeable (on the scale of the figure) renormalization induced by the light--matter coupling. Such low-energy plasmonic modes are thus coined ``dark'' modes since they only weakly couple to light. 
They correspond to LSPs within each sublattices that are out of phase. 
Conversely, plasmonic modes which show a significant coupling to light are called ``bright'' modes.
Such modes correspond to LSPs within each sublattices that are in phase.
Notably, the different nature of the bright and dark modes has also important consequences on their respective radiative lifetimes, as we will discuss in  
Sec.~\ref{sec:rad_linewidths}.

Importantly, the Dirac cone exhibited by the quasistatic band structure remains unaffected by the light--matter coupling since the Dirac point lays outside of the light cone within the regime of parameters which we consider in this work. In Fig.~\ref{fig.Disp_retarded}(b), we nevertheless observe a slight mismatch between the two bands in the vicinity of the $K$ point. Such an artifact stems from the fact that we only consider the light cone belonging to the 1BZ in the evaluation of Eq.~\eqref{eq:rad_shift_general}. Full polaritonic numerical calculations~\cite{Mann-NatCom2018}, where the light--matter interaction is taken into account exactly (and not perturbatively), show however that Dirac cones are unaffected by retardation effects.

We show in Fig.\ \ref{fig.Disp_retarded}(c) the dispersion relation 
including retardation effects of the OP plasmonic modes 
for the Lieb lattice. Along the lines of the above discussion on the honeycomb lattice [cf.\ Fig.~\ref{fig.Disp_retarded}(b)], the plasmonic dispersion relation of the Lieb lattice present bright and dark modes. The most energetic band (red line) corresponds to bright transverse collective modes and thus displays a singularity at the intersection between the quasistatic dispersion (gray solid lines) and the light cone (dashed lines) as well as the absence of a cusp at the $\Gamma$ point. Conversely, the two low-energy bands correspond to dark modes. 
 
We now focus on the effects of retardation in the light--matter interaction onto the collective plasmonic modes polarized within the plane of the metasurface [cf.\ Eq.~\eqref{eq:delta_inplane}]. We display in Fig.~\ref{fig.Disp_retarded}(d) the dispersion relation of the IP modes in the square lattice (see red and orange lines). The quasistatic band structure from Fig.~\ref{fig.Disp-square-z}(d) is reproduced here by gray lines for comparison. Three important features appear from the comparison of these two results. The low-energy band (cf.\ red line), which corresponds essentially to transverse modes, present a singularity at the crossing of the quasistatic dispersion relation with the light cone. Conversely, the high-energy band (cf.\ orange line), which corresponds essentially to longitudinal modes, does not present such a singularity at the crossing, as is the case for plasmonic chains \cite{Weber-PRB2004, Citrin-Nanolett2004, Simovski-PRE2005, Citrin-OptLett2006, Koenderink-PRB2006, Markel-PRB2007, Fung-OptLett2007, Petrov-PRA2015, downi18_JPCM, downi18_EPJB}. In addition, the cusp 
that present the quasistatic high-energy band in the vicinity of the $\Gamma$ point is washed away by retardation effects. 
Similar conclusions can be drawn for the honeycomb and Lieb lattices, see Figs.~\ref{fig.Disp_retarded}(e) and \ref{fig.Disp_retarded}(f), respectively, 
as well as for the kagome lattice \cite{SM}. 
In addition, in the case of Bravais lattices with a basis, some of the IP bands are only weakly modified by retardation effects 
[see the cyan lines in Figs.~\ref{fig.Disp_retarded}(e) and \ref{fig.Disp_retarded}(f)] and thus correspond to dark modes. 

In Appendix \ref{app:Han}, we compare our perturbative approach to numerical calculations based on the solution to the fully retarded Maxwell's equations performed in Refs.~\cite{Han-PRL2009, zhen08_PRB}, and find a rather satisfying agreement. 
Contrarily to three-dimensional plasmonic systems \cite{lamow18_PRB} or metamaterials embedded in a photonic cavity \cite{downi19_preprint, Mann-NatCom2018, mann20_preprint}, where matter and photonic degrees of freedom are strongly coupled and which require an exact diagonalization of the light-matter coupling Hamiltonian, 
here the collective plasmons supported by the metasurface couple to a continuum of photonic modes as $\mathcal{V}\rightarrow\infty$ in Eq.~\eqref{Vector-potential}. Therefore, the coupling is not as strong as in the previously mentioned systems, and leading-order perturbation theory is able to catch the main features of the polaritonic band structure.

\section{Radiative damping}
\label{sec:rad_linewidths}
We now concentrate on the evaluation of the radiative decay rate of the collective plasmons. To this end, we treat the plasmon--photon coupling Hamiltonian \eqref{H-plph} as a weak perturbation to the plasmonic subsystem. In such a regime, the radiative decay rate of the plasmonic eigenmode $|1_{\tau}^{\varepsilon}(\q)\rangle$ with band index $\tau$, polarization $\varepsilon$, and wave vector $\q$  
is given by the Fermi golden rule expression
\begin{align}
\label{eq:rad_damping_FGR}
\gamma_{\tau}^{\varepsilon}(\q)=&\;2\pi^{2}\omega_{0}^{3}\frac{a^{3}}{\mathcal{V}}\sum_{\mathbf{k},\hat\lambda_\mathbf{k}}
\frac{1}{\omegak}\delta\big(\omegatau-\omegak\big)\nonumber \\ 
&\times\left|F_{\mathbf{k},\q}^{-}\sum_{s\sigma}(\hat{\sigma}\cdot\hat\lambda_\mathbf{k})\mathrm{e}^{-\mathrm{i}(\q-\mathbf{k})\cdot\ds}P_{\tau s}^{\varepsilon\sigma}(\q)\right|^{2},
\end{align}
where $F_{\mathbf{k},\q}^{-}$ and $P_{\tau s}^{\varepsilon\sigma}(\q)$ are defined in Eqs.\ \eqref{eq:array_factor} and \eqref{eq:P}, respectively. In the large metasurface limit ($\mathcal{N}\gg 1$), using Eq.~\eqref{eq:array_factor_infinite} yields the result
\begin{align}
\label{eq:rad_damping_general}
\gamma_{\tau}^{\varepsilon}(\q)=&\;2\pi^{2}\omega_{0}^{3}\frac{a^{3}}{\mathcal{V}}\sum_{\mathbf{k},\hat\lambda_\mathbf{k}}
\frac{|F_{\mathbf{k},\q}^{-}|^{2}}{\omegak}\left|\sum_{s\sigma}(\hat{\sigma}\cdot\hat\lambda_\mathbf{k})P_{\tau s}^{\varepsilon\sigma}(\q)\right|^{2}\nonumber \\ 
&\times\delta\big(\omegatau-\omegak\big),
\end{align}
which is valid for arbitrary polarizations. 
We can then show that \cite{SM}
\begin{align}
\label{eq:gamma_z}
\gamma_{\tau}^{z}(\q)=&\;\frac{2\pi\omega_{0}^{3}a^{3}c|\q|^{2}}{|\ta\times\tb|\left[\omega_{\tau}^{z}(\q)\right]^{2}}\left|\sum_{s}P_{\tau s}^{zz}(\q)\right|^{2}
\nonumber \\ 
&\times\frac{\Theta\big(\omega_\tau^z(\q)-c|\q|\big)}{\sqrt{\left[\omega_\tau^z(\q)\right]^2-\left(c|\q|\right)^2}}
\end{align}
for the OP modes, and
\begin{align}
\label{eq:gamma_inplane}
\gamma_\tau^{\varepsilon_\parallel}(\q)=&\;-\frac{2\pi\omega_0^3a^3 c|\q|^{2}}{|\mathbf{t}_1\times\mathbf{t}_2|
\left[\omega_\tau^{\varepsilon_\parallel}(\q)\right]^2}
\frac{\Theta\big(\omega_\tau^{\varepsilon_\parallel}(\q)-c|\q|\big)}{\sqrt{\left[\omega_\tau^{\varepsilon_\parallel}(\q)\right]^2-\left(c|\q|\right)^2}}
\nonumber\\
&\times
\Bigg\{
\left|\sum_{s}\left[\frac{q_x}{|\q|}P_{\tau s}^{\varepsilon_\parallel x}(\q)
+\frac{q_y}{|\q|}P_{\tau s}^{\varepsilon_\parallel y}(\q)\right]\right|^2
\nonumber\\
&-
\left[
\left|\sum_{s}P_{\tau s}^{\varepsilon_\parallel x}(\q)\right|^2
+
\left|\sum_{s}P_{\tau s}^{\varepsilon_\parallel y}(\q)\right|^2
\right]
\nonumber\\
&\times
\frac{\left[\omega_\tau^{\varepsilon_\parallel}(\q)\right]^{2}}
{\left(c|\q|\right)^2}
\Bigg\}
.
\end{align}
for the IP modes.

\begin{figure*}[t!]
\begin{center}
\includegraphics[width=\textwidth]{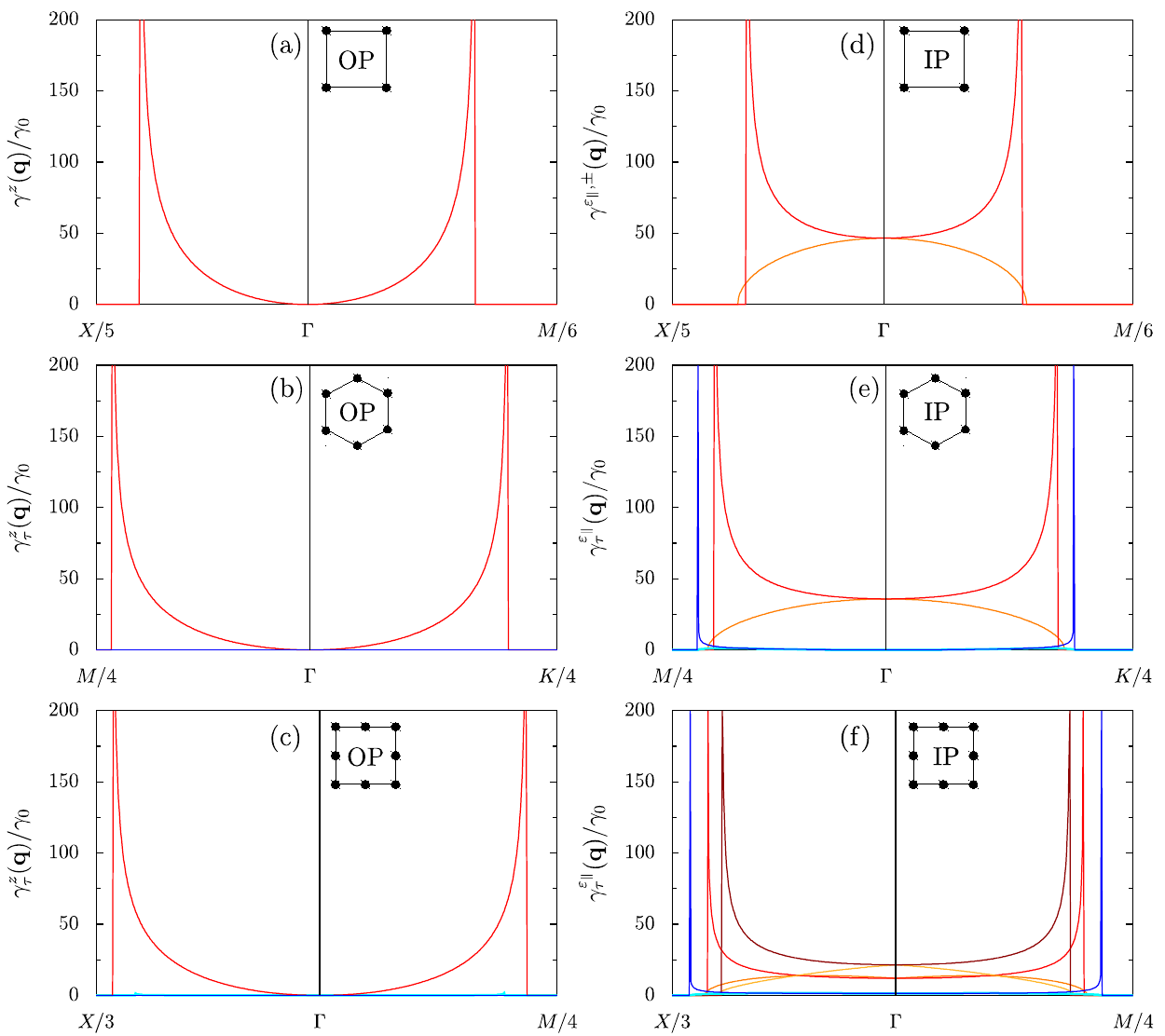}
\caption{\label{fig.radiative_damping} The colored lines display the radiative damping rate $\gamma_{\tau}^{\varepsilon}(\q)$ from Eq.~\eqref{eq:rad_damping_general} [scaled by the single-particle result $\gamma_{0}$, see Eq.~\eqref{eq:gamma_0}] for (a) and (d) the square, (b) and (e) the honeycomb, and (c) and (f) the Lieb lattices. (a)--(c) [(d)--(f)] The damping for the OP [IP] modes. Same parameters as in Fig.\ \ref{fig.Disp_retarded}.}\end{center}
\end{figure*}

We plot in Fig.~\ref{fig.radiative_damping} the radiative damping rate \eqref{eq:gamma_z} for OP plasmonic modes for the square 
[Fig.~\ref{fig.radiative_damping}(a)], the honeycomb [Fig.~\ref{fig.radiative_damping}(b)], and the Lieb lattices [Fig.~\ref{fig.radiative_damping}(c)] along the high symmetry lines originating from the $\Gamma$ point of their respective 1BZ. Note that we do not show in the figure the results along the entire 1BZ since the damping rate vanishes for wave vectors outside of the light cone, as can be easily inferred from the Heaviside step function in Eq.~\eqref{eq:gamma_z}. In Fig.\ \ref{fig.radiative_damping} the displayed results are scaled by the radiative decay rate of a single isolated nanoparticle, 
\begin{equation}
\label{eq:gamma_0}
\gamma_{0}=\frac{2\omega_{0}^{4}a^{3}}{3c^{3}}.
\end{equation}
 For the square lattice [Fig.~\ref{fig.radiative_damping}(a)], the OP plasmonic modes present a highly superradiant profile ($\gamma^{z}(\q)\gg\gamma_{0}$) inside of the light cone for wave vectors not too close to the $\Gamma$ point, while displaying a vanishing rate outside of the light cone. From the figure we observe that the radiative decay rate $\gamma^{z}(\q)$ increases rapidly as $\q$ moves away from the center of the 1BZ and diverges at wave vectors corresponding to the intersection of the quasistatic band structure with the light cone. 
Such singularities are related to those observed in the fully retarded dispersion relation [cf.\ Fig.~\ref{fig.Disp_retarded}(a)].
The superradiant behavior of the radiative damping rate observed in Fig.~\ref{fig.radiative_damping}(a) is reminiscent to the one reported for 1D plasmonic chains~\cite{Weber-PRB2004, Citrin-Nanolett2004, Markel-PRB2007, Citrin-OptLett2006, Koenderink-PRB2006, Petrov-PRA2015, Brandstetter-PRB2016, Downing-PRB2017, downi18_JPCM, downi18_EPJB}. Such a behavior is nevertheless much more prominent in 2D metamaterials, due to the enhanced constructive interferences between the dipolar electric fields produced by each LSP. 

We plot in Figs.\ \ref{fig.radiative_damping}(b) and \ref{fig.radiative_damping}(c) the radiative decay rates of the OP plasmonic modes for the honeycomb and Lieb lattices, respectively. In both cases, one of the plasmonic band shows a similar profile as that in Fig.~\ref{fig.radiative_damping}(a). Indeed, in both Figs.~\ref{fig.radiative_damping}(b) and \ref{fig.radiative_damping}(c), the red lines display a superradiant behavior which diverges at the intersection of the quasistatic band structure and the light cone. Conversely, the blue and cyan lines show the existence of subradiant modes (i.e., $\gamma_{\tau}^{\varepsilon}(\q)\ll \gamma_{0}$) corresponding to dark, nonradiative modes, for which retardation effects on the plasmonic band 
structure are essentially negligible [cf.\ blue and cyan lines in Figs.~\ref{fig.Disp_retarded}(b) and \ref{fig.Disp_retarded}(c)].

We show in Fig.\ \ref{fig.radiative_damping}(d) the radiative damping rate \eqref{eq:gamma_inplane} for IP polarized modes in the square lattice. In the figure, the red line corresponds to the lower transverse plasmonic band in 
Fig.~\ref{fig.Disp_retarded}(d) [see also Fig.~\ref{fig.Disp-square-z}(d)] and presents singularities coinciding with those in Fig.~\ref{fig.radiative_damping}(a). Conversely, the orange line in Fig.\ \ref{fig.radiative_damping}(d) which corresponds to the upper longitudinal band [see Fig.\ \ref{fig.Disp_retarded}(d)] displays an opposite trend, as the radiative decay rate decreases for wave vectors moving away from the center of the 1BZ. This is reminiscent to what has been previously reported for longitudinal plasmonic modes in 1D chains~\cite{Weber-PRB2004, Citrin-Nanolett2004, Markel-PRB2007, Citrin-OptLett2006, Koenderink-PRB2006, Petrov-PRA2015, Brandstetter-PRB2016, Downing-PRB2017, downi18_JPCM, downi18_EPJB}. 
We draw similar conclusions for the honeycomb and Lieb lattices, see Figs.~\ref{fig.radiative_damping}(e) and \ref{fig.radiative_damping}(f), respectively. Additionally, some of the bands [cyan lines in Figs.\ \ref{fig.radiative_damping}(e) and \ref{fig.radiative_damping}(f)] display an almost vanishing radiative decay rate as they correspond to dark, out-of-phase modes.

\section{Effects of the electronic environment on the collective plasmon excitations}
\label{sec:Landau}

In this section we now focus on the effects induced by the second environment 
the collective plasmons are subject to [cf. Eq.~\eqref{System-Hamiltonian}], i.e., 
the electronic environment, which 
is represented by the Hamiltonian \eqref{H-eh}. 
Similarly to the photonic bath which we considered in the preceding Secs.~\ref{sec:rad_frequency_shifts} and \ref{sec:rad_linewidths}, the coupling between plasmonic and single-particle electronic degrees of freedom, encapsulated in the Hamiltonian \eqref{H-pleh}, leads to two distinct effects. First, 
the collective plasmons dissipate their energy by producing electron-hole pairs inside each nanoparticles composing the metasurface, corresponding 
to the well-known Landau damping (Sec.~\ref{sec:Landau_linewidth}), and yielding a second (nonradiative) 
decay channel which adds 
up to the radiative one. Second, the electronic environment induces an additional renormalization of the quasistatic dispersion relation, which comes on top of the one induced by free photons 
(Sec.~\ref{sec:Landau_shift}).

\subsection{Landau damping}
\label{sec:Landau_linewidth}
We start this section by first evaluating the Landau damping of the collective plasmonic modes. 
To this end, we treat the coupling \eqref{H-pleh} between plasmonic and electronic degrees of freedom perturbatively. 
Since the typical Fermi temperature of ordinary metals is of the order of $\unit[10^4]{K}$, we employ the zero-temperature limit which is a very good approximation. 
Within this regime, the Landau damping linewidth $\Gamma_{\tau}^{\varepsilon}(\q)$ of the plasmonic eigenmode $|1_\tau^\varepsilon(\q)\rangle$ with band index $\tau$, polarization $\varepsilon$, and wave vector $\q$ is given by the Fermi golden rule 
\begin{align}
\label{eq:damping_Landau}
\Gamma_{\tau}^{\varepsilon}(\q)=&\;\frac{2\pi}{\hbar^2}\Lambda^2
\sum_s \sum_{eh}\left\vert\sum_\sigma [M_{\tau s}^{\varepsilon\sigma}(\q)]^* 
\langle e|\sigma|h\rangle\right\vert^2 \nonumber \\
& \times\delta\big(\omegatau-\omega_{eh}\big), 
\end{align}
where 
\begin{equation}
\label{eq:M}
M_{\tau s}^{\varepsilon\sigma}(\q)=u_{\tau s}^{\varepsilon\sigma}(\q)-v_{\tau s}^{\varepsilon\sigma}(\q)
\end{equation} 
is given in terms of the Bogoliubov coefficients entering Eq.~\eqref{Beta-operators}.
In Eq.~\eqref{eq:damping_Landau}, $\Lambda$ is the coupling constant defined in Eq.~\eqref{eq:Lambda}, $\omega_{eh}=(\epsilon_e-\epsilon_h)/\hbar$ corresponds to the frequency associated to an electron-hole pair, where $\epsilon_e$ ($\epsilon_h$) is the energy of a single-particle electron (hole) state in the self-consistent hard-wall potential associated to each nanoparticle, and $\langle e|\sigma|h\rangle$ 
is the corresponding dipolar matrix element \cite{SM}.

Equation~\eqref{eq:damping_Landau} can be explicitly calculated \cite{SM}, 
yielding for the Laudau damping rate of the collective plasmons the analytical expression
\begin{align}
\label{eq:gamma_Landau}
\Gamma_{\tau}^{\varepsilon}(\q)=\frac{3v_{\mathrm{F}}}{4a}
\left(\frac{\omega_0}{\omegatau}\right)^4 
g\left(\frac{\hbar\omegatau}{E_\mathrm{F}}\right).
\end{align}
Here, $v_{\mathrm{F}}$ and $E_{\mathrm{F}}$ are the Fermi velocity and energy of the considered metal, respectively, while the function $g$ is a is a monotonically decreasing function 
of the parameter $\hbar\omegatau/E_\mathrm{F}$ \cite{SM, Yannouleas-AP1992}.
To leading order in the coupling constant \eqref{eq:Omega}, the Landau damping decay rate of the collective plasmonic modes thus scales with the nanoparticle size as $1/a$, as it is the case for the single-particle result \cite{SM, Yannouleas-AP1992, Kubo-JPSJ1966, Barma1989}. This is in stark contrast to the radiative linewidth, which increases with the nanoparticle radius as $a^3$ [cf.\ Eq.~\eqref{eq:rad_damping_general}].
 
\begin{figure}[t!]
\begin{center}
\includegraphics[width=\columnwidth]{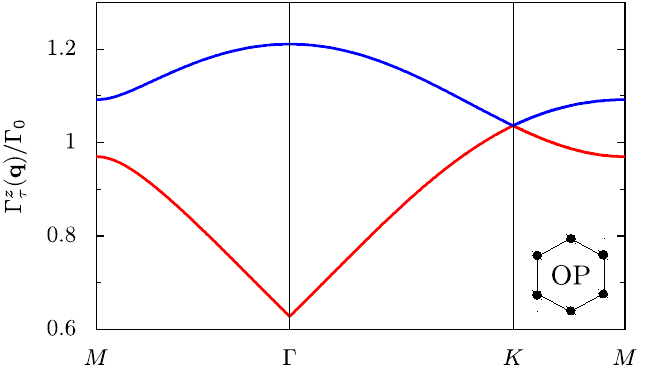}
\caption{\label{fig:Damping_Landau}Landau damping linewidth \eqref{eq:gamma_Landau} (scaled by $\Gamma_0$, the Landau decay rate of a single nanoparticle) of the out-of-plane polarized modes in the honeycomb lattice. The blue (red) line corresponds to the decay rate of the lower (higher) band [see Fig.\ \ref{fig.Disp-honeycomb-z}(c)]. In the figure, $d=3a$ and $E_{\mathrm{F}}=2\hbar\omega_0$.}
\end{center}
\end{figure}

In Fig.\ \ref{fig:Damping_Landau}, we illustrate the result of Eq.\ \eqref{eq:gamma_Landau} for the special case of the OP modes in the honeycomb lattice. In the figure, the displayed results are scaled by the single-particle Landau damping linewidth 
$\Gamma_0=(3 v_\mathrm{F}/4a) g(\hbar\omega_0/E_\mathrm{F})$ \cite{SM}, and 
the blue (red) curve corresponds to the lower (higher) energy band in Fig.~\ref{fig.Disp-honeycomb-z}(c).
We therefore conclude that the higher the energy of the mode, the lower its nonradiative Landau decay rate, as 
is the case for single nanoparticles \cite{Seoanez-EPJD2007}. Importantly, in contrast to the radiative decay rate analyzed in Sec.~\ref{sec:rad_linewidths}, 
the Landau damping is nonvanishing over the whole 1BZ, and is of the order of $\Gamma_0$. It is therefore crucial to take into account such a decay mechanism, especially for small nanoparticles where it dominates over radiative losses~\cite{SM}.

\subsection{Electronic-induced frequency shift}
\label{sec:Landau_shift}

We now calculate analytically the frequency shift induced by the electronic environment on the plasmonic band structure. Treating the plasmon-electron coupling Hamiltonian \eqref{H-pleh} to second order in perturbation theory yields for the collective plasmon energy levels 
$n_\tau^\varepsilon(\q)\hbar\omegatau
+\mathcal{E}_{n_\tau^\varepsilon(\q)}^{(1)}
+\mathcal{E}_{n_\tau^\varepsilon(\q)}^{(2)}$. While the first-order correction $\mathcal{E}_{n_\tau^\varepsilon(\q)}^{(1)}$ vanishes due to the selection rules contained in the coupling Hamiltonian \eqref{H-pleh}, the zero-temperature second-order correction reads 
 \begin{align}
\label{eq:E_elec_2}
\mathcal{E}_{n_\tau^\varepsilon(\q)}^{(2)}=&\; \frac{\Lambda^2}{\hbar}
\sum_s  \sum_{eh}\left[\frac{n_\tau^\varepsilon(\q)}
{\omegatau-\omega_{eh}}
\left|\sum_\sigma
[M_{\tau s}^{\varepsilon\sigma}(\q)]^* \langle e|\sigma|h\rangle\right|^2
\right.
\nonumber\\
&-\left.\frac{n_\tau^\varepsilon(\q)+1}
{\omegatau+\omega_{eh}}
\left|\sum_\sigma
M_{\tau s}^{\varepsilon\sigma}(\q) \langle e|\sigma|h\rangle\right|^2
\right].
\end{align}
The electronic-induced frequency shift, defined as 
$\Delta_\tau^\varepsilon(\q)=[\mathcal{E}_{n_\tau^\varepsilon(\q)+1}^{(2)}
-\mathcal{E}_{n_\tau^\varepsilon(\q)}^{(2)}]/\hbar$, is therefore given by
\begin{align}
\label{eq:electronic_shift}
\Delta_{\tau}^\varepsilon(\q)=&\; \frac{\Lambda^2}{\hbar^2}
\sum_s \sum_{eh}\left[\frac{1}
{\omegatau-\omega_{eh}}\left|\sum_\sigma
[M_{\tau s}^{\varepsilon\sigma}(\q)]^* \langle e|\sigma|h\rangle\right|^2
\right.
\nonumber\\
&-\left.\frac{1}
{\omegatau+\omega_{eh}}\left|\sum_\sigma 
M_{\tau s}^{\varepsilon\sigma}(\q) \langle e|\sigma|h\rangle\right|^2
\right].
\end{align}

The above expression can be evaluated analytically~\cite{SM}, leading to the result
\begin{equation}
\label{eq:Landau_shift_final}
\Delta_\tau^\varepsilon(\q)=-\frac{3v_{\mathrm{F}}}{4a}\beth\big(\omegatau\big),
\end{equation}
where
\begin{align}
\label{eq:beth_function}
\beth(\omega)=&\;\frac{4}{15\pi}\sqrt{\frac{E_{\mathrm{F}}}{\hbar\omega}}
\left(
\frac{\omega_0}{\omega}\right)^4
\nonumber\\
&\times\Bigg[
\ln\left(
\frac{8\hbar\omega k_{\mathrm{F}}a}{3\eta E_{\mathrm{F}}g(
\hbar\omega/E_{\mathrm{F}})}\left[\frac{\omega}{\omega_0}\right]^4
\right)
-\frac{\pi}{2}-\frac{4}{3}
\Bigg].
\end{align}
Here, $\eta$ is a constant of order $1$ \cite{SM}. 
Note that the frequency shift \eqref{eq:Landau_shift_final} scales with the size of the nanoparticles as $1/a$, up to a logarithmic correction. As is the case for the 
Landau damping linewidth \eqref{eq:gamma_Landau}, such a frequency renormalization is therefore of relevance
for the smallest nanoparticles. Moreover, the electronic shift of the collective plasmons involves only the plasmonic band structure $\omegatau$ in contrast to the radiative frequency shifts [cf.\ Eqs.\ \eqref{eq:delta_z} and \eqref{eq:delta_inplane}] which depend on the eigenvectors as well. Notice that substituting $\omega_0$ with $\omegatau$ in Eq.\ \eqref{eq:Landau_shift_final} allows us to recover the electronic-induced frequency redshift of an isolated nanoparticle \cite{Weick-PRB2006},
\begin{equation}
\label{eq:Landau_shift_single_NP}
\Delta_0=-\frac{3v_{\mathrm{F}}}{4a}\beth(\omega_0).
\end{equation}

\begin{figure}[t!]
\begin{center}
\includegraphics[width=\columnwidth]{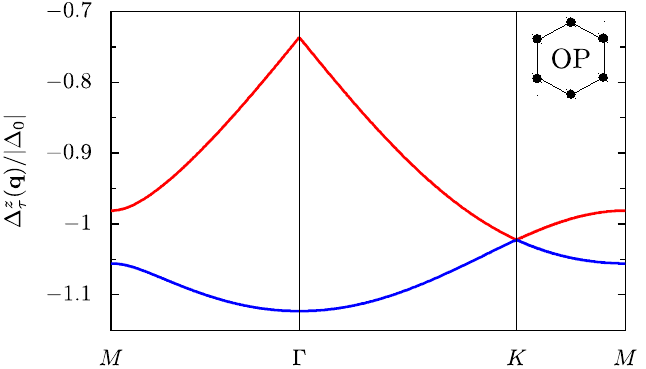}
\caption{\label{fig:shift_Landau}Electronic-induced frequency shift \eqref{eq:Landau_shift_final} [scaled by $|\Delta_0|$, see Eq.~\eqref{eq:Landau_shift_single_NP}] 
of the out-of-plane polarized modes in the honeycomb lattice. The blue (red) line corresponds to the frequency shift associated with the lower (higher) plasmonic band [see Fig.~\ref{fig.Disp-honeycomb-z}(c)]. In the figure, $d=3a$,
$E_{\mathrm{F}}=2\hbar\omega_0$, $k_{\mathrm{F}}a=100$, and $\eta=1/2$.}
\end{center}
\end{figure}

We plot in Fig.~\ref{fig:shift_Landau} the result of Eq.~\eqref{eq:Landau_shift_final} applied to the special case of the OP collective plasmonic modes in the honeycomb lattice. As in Fig.\ \ref{fig:Damping_Landau}, the blue (red) line in Fig.\ \ref{fig:shift_Landau} corresponds to the lower (higher) energy band in Fig.\ \ref{fig.Disp-honeycomb-z}(c). The displayed results are scaled by the absolute value of the frequency shift \eqref{eq:Landau_shift_single_NP} corresponding to a single nanoparticle. Immediately noticeable from the figure is that the electronic shift \eqref{eq:Landau_shift_final} is negative in the entire 1BZ, thus corresponding to a redshift, and is of the order of $\Delta_0$.
This is in contrast to the radiative frequency shift \eqref{eq:rad_shift_general}, whose sign is both depending on the wave vector and polarization of the collective mode (see Sec.~\ref{sec:rad_frequency_shifts}). Finally, the higher the energy of the mode, the lower is its associated electronic shift. Such a conclusion is reminiscent of what occurs in isolated nanoparticles~\cite{Seoanez-EPJD2007}.

\section{Observability of the collective plasmonic modes}
\label{sec:exp}

Experimentally, the ability to observe the plasmonic dispersion relations presented, e.g., in Fig.\ \ref{fig.Disp_retarded}, is governed by the resolution of the separation between the bands with respect to their respective linewidths. 
The spectral function $\mathcal{A}(\omega, \q)$, which characterizes the response of the system to an external perturbation at a given frequency $\omega$ and in-plane wave vector $\q$, is the key quantity which is usually determined in a spectroscopy experiment (using, e.g., photons or hot electrons). 
Assuming a Breit--Wigner form for $\mathcal{A}(\omega, \q)$, 
we have
\begin{subequations}
\label{eq:Absorption}
\begin{equation}
\label{eq:Absorption_OP}
\mathcal{A}_{\mathrm{OP}}(\omega,\q)\propto\sum_{\tau}
\frac{1}
{\left[\omega-\tilde{\omega}_\tau^z(\q)\right]^2
+\left[\daleth_\tau^z(\q)/2\right]^2}
\end{equation}
for the OP modes and
\begin{equation}
\label{eq:Absorption_IP}
\mathcal{A}_{\mathrm{IP}}(\omega,\q)\propto\sum_{\tau,\varepsilon_{\parallel}}
\frac{1}
{\left[\omega-\tilde{\omega}_\tau^{\varepsilon_{\parallel}}(\q)\right]^2
+\left[\daleth_\tau^{\varepsilon_{\parallel}}(\q)/2\right]^2}
\end{equation}
\end{subequations}
for the IP modes, respectively. 
In Eq.~\eqref{eq:Absorption}, the renormalized resonance frequency 
\begin{equation}
\tilde{\omega}_\tau^{\varepsilon}(\q)=\omega_\tau^{\varepsilon}(\q)+\delta_\tau^{\varepsilon}(\q)+\Delta_\tau^{\varepsilon}(\q)
\end{equation}
 takes into account both the radiative shift $\delta_\tau^{\varepsilon}(\q)$ [cf.\ Eq.~\eqref{eq:rad_shift_general}] and the electronic one $\Delta_\tau^{\varepsilon}(\q)$
[cf.\ Eq.~\eqref{eq:Landau_shift_final}]. Additionally, the quantity 
\begin{equation}
\label{eq:daleth}
\daleth_\tau^\varepsilon(\q)=\gamma_\tau^\varepsilon(\q)
+\Gamma_\tau^\varepsilon(\q)+\gamma_{\mathrm{O}}
\end{equation}
is the total linewidth of the plasmonic modes and includes three distinct contributions: (i) the radiative losses $\gamma_\tau^\varepsilon(\q)$ [Eq.~\eqref{eq:rad_damping_general}], (ii) the Landau damping $\Gamma_\tau^\varepsilon(\q)$ [Eq.~\eqref{eq:damping_Landau}], and (iii) the Ohmic losses, inherent to any (bulk) metal, characterized by the damping rate $\gamma_{\mathrm{O}}$, and which we assume to be mode and size independent.

\begin{figure*}[t!p]
\begin{center}
\includegraphics[width=.84\textwidth]{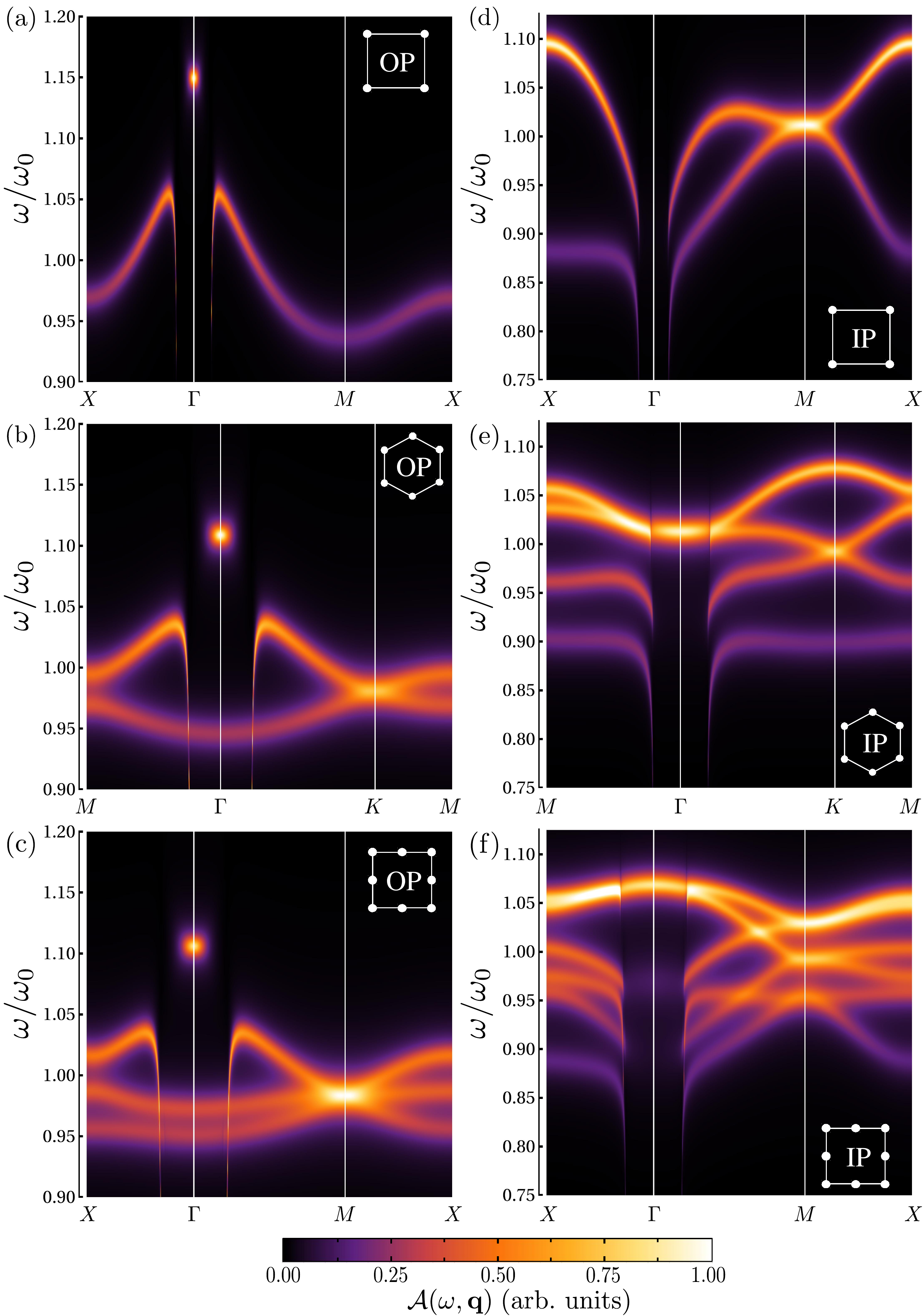}
\caption{\label{fig:Absorption_no_Ohmic} Spectral function \eqref{eq:Absorption} for (a) and (d) the square, (b) and (e) the honeycomb, and (c) and (f) the Lieb lattices. (a)--(c) [(d)--(f)] The results for the OP [IP] modes. In the figure we choose the parameters $\hbar\omega_0/E_{\mathrm{F}}=0.47$ and $k_0/k_{\mathrm{F}}=1.1\times10^{-3}$, corresponding to silver nanoparticles. 
In addition, we take $d=3a$ and $k_0 a =0.15$ (corresponding to $a=\unit[11]{nm}$).  The Ohmic losses $\gamma_{\mathrm{O}}$ entering the total damping rate \eqref{eq:daleth} are neglected.}
\end{center}
\end{figure*}

We show in Fig.~\ref{fig:Absorption_no_Ohmic} the spectral function \eqref{eq:Absorption} for both the OP and IP modes and for the square, the honeycomb, and the Lieb lattices along the high symmetry lines of their respective 1BZ as a function of $\omega$. 
The parameters used in the figure are $\omega_0=\unit[2.6]{eV/\hbar}$  and $E_\mathrm{F}=\unit[5.5]{eV}$, corresponding to Ag nanoparticles placed in an effective medium with a refractive index of $2$ \cite{Kreibig1995}.
The interparticle distance is $d=3a$ and the (reduced) nanoparticle radius is $k_0 a=0.15$ (corresponding to $a=\unit[11]{nm}$).
In Fig.~\ref{fig:Absorption_no_Ohmic} the Ohmic losses entering Eq.~\eqref{eq:daleth} are neglected.

As can be seen in Fig.\ \ref{fig:Absorption_no_Ohmic}(a), which displays $\mathcal{A}(\omega, \q)$ for the OP modes in the square lattice, the spectral function is reminiscent of the dispersion relation including retardation effects shown in Fig.\ \ref{fig.Disp_retarded}(a), 
since the electronic shift \eqref{eq:Landau_shift_final} only induces a finite $\q$-dependent redshift of the band structure. 
In the figure, we clearly distinguish two different profiles corresponding to wave vectors inside ($|\q|\lesssim k_0$) and outside ($|\q|\gtrsim k_0$) of the light cone. 
Within the light cone, the total linewidth \eqref{eq:daleth} is dominated by the radiative damping [see Fig.~\ref{fig.radiative_damping}(a)], so that it is difficult to resolve the plasmonic dispersion relation. However, outside of the light cone, only the Landau damping~\eqref{eq:gamma_Landau}
contributes to the total linewidth of the spectral function, allowing for a clear resolution of the plasmonic band. 
Note that, since the plasmonic modes outside of the light cone are essentially dark, 
nonoptical techniques are required to excite such modes.  
Dark modes in nanoparticle dimers and chains have been observed using electron energy loss spectroscopy 
experiments \cite{Chu-Nanolett2009,Koh-ACSNano2009,Barrow-NanoLett2014}, and such a technique may be transposed to study plasmonic metasurfaces.

We plot in Figs.\ \ref{fig:Absorption_no_Ohmic}(b) and \ref{fig:Absorption_no_Ohmic}(c) the spectral function \eqref{eq:Absorption_OP} corresponding to the OP modes for the honeycomb and Lieb lattices, respectively. In both cases, we observe a similar trend as that in Fig.\ \ref{fig:Absorption_no_Ohmic}(a). Indeed, the entire band structure is clearly visible outside of the light cone in both Figs.~\ref{fig:Absorption_no_Ohmic}(b) and \ref{fig:Absorption_no_Ohmic}(c), while the upper bands inside the light cone display a rather large linewidth. However, we see that the lower bands for both the honeycomb and the Lieb lattices are still well resolved inside of the light cone. These dispersion relations correspond to dark plasmonic modes, so that their radiative linewidths for wave vectors $|\q|\lesssim k_0$ are nearly vanishing [see the blue line in Fig.~\ref{fig.radiative_damping}(b) and the blue and cyan lines in Fig.~\ref{fig.radiative_damping}(c)], and only the Landau damping contributes to the observed linewidth.

In Figs.\ \ref{fig:Absorption_no_Ohmic}(d)--\ref{fig:Absorption_no_Ohmic}(f) we display the spectral function \eqref{eq:Absorption_IP} for the IP plasmonic modes  for (d) the square, (e) the honeycomb, and (f) the Lieb lattices, respectively. Similar conclusions
as that drawn above for the OP modes can be put forward: while the nonradiative bands outside of the light cone are clearly visible, the bright ones
inside of the light cone are essentially almost not resolvable.

We conclude this section by discussing the effect of Ohmic damping on the above results. 
As is evident from Eqs.~\eqref{eq:Absorption} and \eqref{eq:daleth}, the inclusion of a finite $\gamma_\mathrm{O}$ leads to a global, size- and mode-independent increase of the width of the spectral lines shown in Fig.~\ref{fig:Absorption_no_Ohmic}. 
From the experiments of Ref.~\cite{Charle1989} on silver clusters, we can extract the value 
$\gamma_\mathrm{O}=\unit[70]{meV/\hbar}$, which corresponds to $0.027\omega_0$.
The increased linewidth induced by Ohmic losses results in a global resolution of the spectral function which is then significantly lower than the one shown in Fig.\ \ref{fig:Absorption_no_Ohmic}. As a consequence, several plasmonic bands cannot be distinguished properly as it is the case, e.g., for the OP modes in the honeycomb and Lieb lattices in the vicinity of the corners of their respective 1BZ \cite{SM}. Nevertheless, within the regime of parameters
used in Fig.~\ref{fig:Absorption_no_Ohmic}, the linewidth of a majority of the plasmonic modes is still sufficiently small to allow for an experimental detection (except for the bright modes within the light cone). 
Embedding the metasurface in a gain media material \cite{Lawandy2004, Citrin-OptLett2006, Maier-2007, Noginov2007, Noginov2009} 
should help diminish the effects induced by 
Ohmic losses on the spectral function and further improve the experimental observability of the collective plasmons.

\section{Conclusion}
\label{sec:conclusion}
We have considered the plasmonic properties of metasurfaces constituted by an ordered arbitrary two-dimensional array of spherical metallic nanoparticles. We have focused on the case where the interparticle distance is much smaller than the wavelength associated with the dipolar localized surface plasmon resonance frequency of single nanoparticles, where the near-field, quasistatic dipole--dipole interaction dominates. We have developed a comprehensive open quantum system framework to analyze in full analytical detail the dispersion relations and the lifetimes of the resulting collective 
plasmonic modes supported by the various metasurfaces which we have studied, including, e.g., the honeycomb and Lieb lattices. Such metasurfaces present appealing topological features in their band structures, such as massless Dirac-like conical dispersions, as well as nearly flat bands, and may be a possible experimental platform to explore new states of hybrid light--matter waves. 

Our model enabled us to unveil analytical expressions for the retarded dispersion relations of the plasmonic collective modes, which are in good agreement with existing numerical solutions to Maxwell's equations. Our model
also includes the effects of the particle-hole environment to which such modes are coupled to, and that are of particular relevance for nanoparticles of only a few nanometers in size. Our theory further allowed us to provide analytical expressions for the radiative and nonradiative (Landau) damping rates of the plasmonic modes, which enabled us to
critically examine their experimental observability. While Ohmic losses, inherent to the metallic nature of plasmonic particles, may make the detection of the collective modes elusive, the use of gain materials should give scope to 
their experimental observation.

\begin{acknowledgments}
We thank C.\ A.\ Downing, D.\ Felbacq, K.\ R.\ Fratus, F.\ Gautier, Y.\ Henry, E.\ Mariani, P.\ I.\ Tamborenea, and D.\ Weinmann for helpful discussions. We acknowledge financial support from Agence Nationale de la Recherche (Grant No.\ ANR-14-CE26-0005 Q-MetaMat).
This research project has been supported by the University of Strasbourg IdEx program.
\end{acknowledgments}

\appendix
\section{Diagonalization of the plasmonic Hamiltonian}
\label{app:diag}

In this Appendix, we present the details of the diagonalization procedure of the plasmonic Hamiltonian \eqref{eq:H_compa}.
Since we consider large metasurfaces, we use periodic boundary conditions and 
move to wave vector space using the Fourier transform 
\begin{equation}
\label{eq:b_Fourier}
b_{s}^{\sigma}(\q)=\frac{1}{\sqrt{\mathcal{N}}}\sum_{\Rs} \mathrm{e}^{-\mathrm{i}\q\cdot\Rs}\, b_{s}^{\sigma}(\Rs)
\end{equation} 
of the bosonic ladder operator $b_{s}^{\sigma}(\Rs)$ defined in Eq.\ \eqref{eq:quantization},
with $\mathcal{N}=\mathcal{N}_{1}\mathcal{N}_{2}\gg1$ the total number of unit cells in the lattice.
The Hamiltonian 
\eqref{eq:H_compa} then reads 
\begin{align}
\Hpl=&\;\hbar\omega_{0}\sum_{\q}\sum_{s}\sum_{\sigma}\bsigdag(\q)\bsig(\q)\nonumber\\
&+\frac{\hbar\Omega}{2}\sum_{\q}\sum_{ss'}\sum_{\sigma\sigma'}\bigg\{\fsig\bsigdag(\q)\nonumber\\
&\times\left[b_{s'}^{\sigma'}(\q)+b_{s'}^{\sigma'\dagger}(-\q)\right]+\mathrm{h.c.}\bigg\}, 
\label{Plasmonic-H}
\end{align}
where $\fsig$ is the lattice sum defined in Eq.~\eqref{fsig}. Note that for $s=s'$, $f_{ss'}^{\sigma\sigma'}(\q)$ is real due to the inversion symmetry of the Bravais lattice. Moreover, $f_{ss'}^{\sigma\sigma'}(\q)=[f_{s's}^{\sigma\sigma'}(\q)]^{*}$ since $\boldsymbol{\rho}_{s's}=-\rhos$. 

The  plasmonic Hamiltonian $\Hpl$ given in Eq.~\eqref{Plasmonic-H} is quadratic, and can thus be diagonalized 
exactly by means of the bosonic Bogoliubov transformation given in Eq.~\eqref{Beta-operators}.
The inverse transformation reads
\begin{equation}
\label{eq:Bogo_inverse}
\bsig(\q)=\sum_{\tau\varepsilon}\left[u_{\tau s}^{\varepsilon\sigma *}(\q)\betatau(\q)-v_{\tau s}^{\varepsilon\sigma *}(\q)\betataudag(-\q)\right].
\end{equation}
Note that the bosonic commutation relations $[\betatau(\q),\betataudagp(\q')]=\delta_{\tau \tau'}\delta_{\varepsilon\varepsilon'}\delta_{\q\q'}$ and $[\betatau(\q),\betataup(\q')]=0$
impose that the Bogoliubov coefficients in Eq.~\eqref{Beta-operators} fulfill the relations
\begin{equation}
\sum_{s\sigma}[\ucoeff u_{\tau' s}^{\varepsilon'\sigma *}(\q)-\vcoeff v_{\tau' s}^{\varepsilon'\sigma *}(\q)]=\delta_{\tau \tau'}\delta_{\varepsilon\varepsilon'}
\end{equation}
and
\begin{equation}
\label{eq:commutator}
\sum_{s\sigma}[\ucoeff v_{\tau' s}^{\varepsilon'\sigma *}(\q)-\vcoeff u_{\tau' s}^{\varepsilon'\sigma *}(\q)]=0,
\end{equation}
where $u_{\tau s}^{\varepsilon\sigma *}(-\q)=\ucoeff$ and $v_{\tau s}^{\varepsilon\sigma *}(-\q)=\vcoeff$. 
The dispersion relation $\omegatau$, as well as the coefficients of the Bogoliubov transformation \eqref{Beta-operators}, are then obtained from the Heisenberg equation of motion 
\begin{equation}
\label{eq:Heisenberg}
\left[\betatau(\q),\Hpl\right]=\hbar\omegatau\betatau(\q),
\end{equation}
which yields the system of equations \eqref{Hopfield-coeff}.

In the case of a generic Bravais lattice (see Sec.~\ref{sec:Bravais}), the diagonalization procedure above yields the dispersion relations \eqref{equ.omega-zz-simple} and \eqref{equ.omega-xy-simple} for the OP and IP modes, respectively. We find 
for the corresponding coefficients of the Bogoliubov transformation~\eqref{Beta-operators} 
\begin{equation}
\label{eq:u_zz}
u^{zz}(\q)= \frac{\omega^{z}(\q)+\omega_{0}}{2\sqrt{\omega_{0}\omega^{z}(\q)}}
\end{equation}
and
\begin{equation}
v^{zz}(\q)= \frac{\omega^{z}(\q)-\omega_{0}}{2\sqrt{\omega_{0}\omega^{z}(\q)}}
\end{equation}
in the case of the OP polarized modes.
For the IP modes, the condition 
$\zeta_\pm^x(\q)\zeta_\pm^y(\q)
=\left[2\omega_0\Omega f^{xy}(\q)\right]^2$ must be fulfilled,
with $\zeta_\pm^\sigma(\q)=[\omega^{\varepsilon_{\parallel,\pm}}(\q)]^2-\omega_0^2-2\omega_0\Omega f^{\sigma\sigma}(\q)$ ($\sigma=x,y$), so that we obtain
\begin{subequations}
\label{eq:u_simpleBravais}
\begin{align}
u^{\varepsilon_{\parallel,\pm} x}(\q)&=\frac{
\omega^{\varepsilon_{\parallel,\pm}}(\q)+\omega_{0}}
{2\sqrt{\omega_{0}\omega^{\varepsilon_{\parallel,\pm}}(\q)}}
\sqrt{\frac{\zeta_\pm^y(\q)}
{\sum_{\sigma=x,y}\zeta_\pm^\sigma(\q)}},
\\
u^{\varepsilon_{\parallel,\pm} y}(\q)&=\pm\mathrm{sgn}\left\{f^{xy}(\q)\right\}
\sqrt{\frac{\zeta_\pm^x(\q)}
{\zeta_\pm^y(\q)}}
u^{\varepsilon_{\parallel,\pm} x}(\q),
\end{align}
\end{subequations}
and
\begin{subequations}
\begin{align}
v^{\varepsilon_{\parallel,\pm} x}(\q)&=\frac{
\omega^{\varepsilon_{\parallel,\pm}}(\q)-\omega_{0}}
{2\sqrt{\omega_{0}\omega^{\varepsilon_{\parallel,\pm}}(\q)}}
\sqrt{\frac{\zeta_\pm^y(\q)}
{\sum_{\sigma=x,y}\zeta_\pm^\sigma(\q)}},
\\
v^{\varepsilon_{\parallel,\pm} y}(\q)&=\pm \mathrm{sgn}\left\{f^{xy}(\q)\right\}
\sqrt{\frac{\zeta_\pm^x(\q)}
{\zeta_\pm^y(\q)}}
v^{\varepsilon_{\parallel,\pm} x}(\q).
\end{align}
\end{subequations}

We point out that neglecting the nonresonant terms in Eq.~\eqref{Plasmonic-H}, which corresponds to performing the rotating wave approximation (RWA), yields the same dispersion relations \eqref{equ.omega-zz-simple} and \eqref{equ.omega-xy-simple}  to first order in $\Omega\ll\omega_0$ and the same Bogoliubov coefficients \eqref{eq:u_zz} and \eqref{eq:u_simpleBravais}, while $v^{\varepsilon\sigma}(\q)=0$ within such an approximation \cite{SM}.

In the case of a generic bipartite lattice (see Sec.~\ref{sec:bipartite}), the Bogoliubov coefficients corresponding to the OP modes are given by 
\begin{subequations}
\label{eq:coeff_bipartite_u}
\begin{align}
u_{\tau 1}^{zz}(\q)&= \frac{\omega_\tau^{z}(\q)+\omega_{0}}{2\sqrt{2\omega_{0}\omega_\tau^{z}(\q)}},
\\
u_{\tau 2}^{zz}(\q)&= \tau\ \frac{f_{12}^{zz}(\q)}{|f_{12}^{zz}(\q)|} u_{\tau 1}^{zz}(\q),
\end{align}
\end{subequations}
and
\begin{subequations}
\label{eq:coeff_bipartite_v}
\begin{align}
v_{\tau 1}^{zz}(\q)&= \frac{\omega_\tau^{z}(\q)-\omega_{0}}{2\sqrt{2\omega_{0}\omega_\tau^{z}(\q)}},
\\
v_{\tau 2}^{zz}(\q)&= \tau\ \frac{f_{12}^{zz}(\q)}{|f_{12}^{zz}(\q)|}v_{\tau 1}^{zz}(\q).
\end{align}
\end{subequations}

Alternatively to the exact diagonalization procedure presented in this Appendix, it may be useful to treat the nonresonant terms in Eq.\ \eqref{Plasmonic-H} perturbatively, since, for all practical purposes, the coupling constant $\Omega\ll\omega_0$ [cf.\ Eq.~\eqref{eq:Omega}]. Such a procedure has the advantage of dividing by two the dimension of the system of equations leading to the plasmonic band structure, which may be helpful in deriving the spectrum and the associated Bogoliubov coefficients analytically \cite{SM}.

\section{Mean-field approximation for the quasistatic band structure of the square lattice}
\label{app:MF}

Here, we provide an analytical understanding of the cusp presented by the quasistatic 
collective plasmon dispersion close to the $\Gamma$ point in the case of the OP and IP polarized modes in the case of the square lattice. As demonstrated in Sec.~\ref{sec:rad_frequency_shifts}, such a cusp is actually an artifact of the quasistatic approximation, and retardation effects strikingly renormalize this behavior. 
It is however important to get a full description of the quasistatic band structure in order to understand the fully retarded one. The analysis presented below, which relies on the mean-field treatment of the long-ranged quasistatic dipolar interactions, can be straightforwardly extended to other Bravais and non-Bravais lattices. 

The behavior of the quasistatic plasmonic dispersion relation of the square lattice for OP polarization close to the $\Gamma$ point mentioned in the main text (see Sec.~\ref{sec:Bravais}) can be understood by treating the nearest neighbors in the lattice sum \eqref{fsig} exactly, while averaging the interactions beyond nearest neighbors in the spirit of the mean-field (mf) approximation, leading to $f^{zz}(\q)\simeq f_{\mathrm{nn}}^{zz}(\q)+f_{\mathrm{mf}}^{zz}(\q)$. Here, 
\begin{align}
f_{\mathrm{mf}}^{zz}(\q)&=\int_{\rho\geqslant\sqrt{2}d}\frac{\mathrm{d}^2\boldsymbol{\rho}}{d^2}\
\mathrm{e}^{\mathrm{i}\q\cdot\boldsymbol{\rho}}\left(\frac{d}{\rho}\right)^3
\nonumber\\
&=2\pi qd\,\mathcal{F}\left(\sqrt{2}|\mathbf{q}|d\right), 
\end{align}
with
\begin{align}
\mathcal{F}\left(z\right)=&\;-1+J_1(z)\left[-1+\frac{\pi}{2}z\, \mathbf{H}_0(z)\right]
\nonumber\\
&+J_0(z)\left[\frac 1z+z-\frac{\pi}{2}z\, \mathbf{H}_1(z)\right], 
\end{align}
where $J_n(z)$ and $\mathbf{H}_n(z)$ are the Bessel functions of the first kind and the Struve functions, respectively. Consequently, in the vicinity 
of the $\Gamma$ point ($|\mathbf{q}|d\ll1$) and in the weak-coupling regime of interest ($\Omega\ll\omega_0$), we find 
\begin{equation}
\label{eq:Bethe-Peierls_square}
\omega^{z}(\q)\simeq\omega_0+\Omega 
\left[4+\sqrt{2}\pi-2\pi |\mathbf{q}|d+\left(\frac{\pi}{\sqrt{2}}-1\right)\left(|\mathbf{q}|d\right)^2
\right],
\end{equation}
so that a cusp appears in the dispersion relation \eqref{equ.omega-zz-simple}. In Fig.~\ref{fig.Disp-square-z}(c) we show by a dotted line the plasmonic band structure within the mean-field approximation detailed above. As can be seen from the figure, the mean-field approximation accurately describes the cusp of the full band structure in the vicinity of the $\Gamma$ point, while it tends toward the nearest-neighbor approximation away from the $\Gamma$ point. 

In the case of the IP-polarized modes, the lattice sums in Eq.~\eqref{equ.omega-xy-simple} are replaced within the mean-field approximation presented above by 
$f^{\sigma\sigma'}(\q)\simeq f^{\sigma\sigma'}_\mathrm{nn}(\q)+f^{\sigma\sigma'}_\mathrm{mf}(\q)$ ($\sigma, \sigma'=x, y$), where
\begin{align}
f^{\sigma\sigma}_\mathrm{mf}(\q)=&\;2\pi |\mathbf{q}|d
\left\{
-\left(\frac{q_\sigma}{|\mathbf{q}|}\right)^2\mathcal{F}\left(\sqrt{2}|\mathbf{q}|d\right)
\right.
\nonumber\\
&+\left.\left[\left(\frac{q_\sigma}{|\mathbf{q}|}\right)^2-\frac 12\right]\frac{J_1\left(\sqrt{2}|\mathbf{q}|d\right)}{(|\mathbf{q}|d)^2}
\right\},\quad \sigma=x,y,
\end{align}
and
\begin{equation}
f^{xy}_\mathrm{mf}(\q)=2\pi |\mathbf{q}|d\frac{q_xq_y}{|\mathbf{q}|^2}
\left[
-\mathcal{F}\left(\sqrt{2}|\mathbf{q}|d\right)+\frac{J_1\left(\sqrt{2}|\mathbf{q}|d\right)}{(|\mathbf{q}|d)^2}
\right].
\end{equation}
The resulting band structure is represented in Fig.~\ref{fig.Disp-square-z}(d) by a dotted line, and reproduces quite well the full 
quasistatic dispersion in the vicinity of the $\Gamma$ point. In particular, to first order in $|\mathbf{q}|d\ll1$, we find $\omega^{\varepsilon_{\parallel,+}}(\q)\simeq\omega_0+\Omega(-2-\pi/\sqrt{2}+2\pi |\mathbf{q}|d)$ and
$\omega^{\varepsilon_{\parallel,-}}(\q)\simeq\omega_0-\Omega(2+\pi/\sqrt{2})$, demonstrating the presence (absence) of a cusp for the high- (low-) energy plasmonic branch.

\section{Comparison of the open quantum system approach to classical electrodynamic calculations}
\label{app:Han}

In this Appendix, we compare our analytical results for the retarded plasmonic band structure derived from the perturbative 
open quantum system approach (see Sec.~\ref{sec:rad_frequency_shifts}) to numerical calculations based on the solution to 
Maxwell's equations~\cite{Han-PRL2009, zhen08_PRB}. 

In Ref.~\cite{Han-PRL2009}, Han \textit{et al.}\ used the multiple scattering theory developed in Refs.~\cite{Fung-OptLett2007, zhen08_PRB} which gives a solution to Maxwell's equations including retardation effects in order to numerically extract the plasmonic dispersion of a honeycomb array of silver nanoparticles in vacuum, with resonance frequency 
$\omega_0=\unit[3.5]{eV/\hbar}$, radius $a=\unit[10]{nm}$, and lattice constant $\unit[60]{nm}$, corresponding to an interparticle distance $d=\unit[35]{nm}$. 
The classical calculation by Han \textit{et al.}\ requires as an input the permittivity of the considered metal, which they chose to be of the Drude-type, without Ohmic losses. We note that the use of a nonlocal dielectric function is not a well-defined concept for a finite-size nanoscale system \cite{Raza_2015}, while the quantum model developed here is inherently nonlocal. Moreover, the method of Refs.~\cite{Fung-OptLett2007, zhen08_PRB, Han-PRL2009} is fully numerical and hence is limited to specific choices of material parameters and geometry of the metasurface, and  involves nonconverging sums that need to be regularized. In contrast, our perturbative approach is straightforward to implement and universal, as it allows to consider any metallic nanoparticles arranged in an arbitrary Bravais or non-Bravais lattice. 

\begin{figure}[tb!]
\begin{center}
\includegraphics[width=\columnwidth]{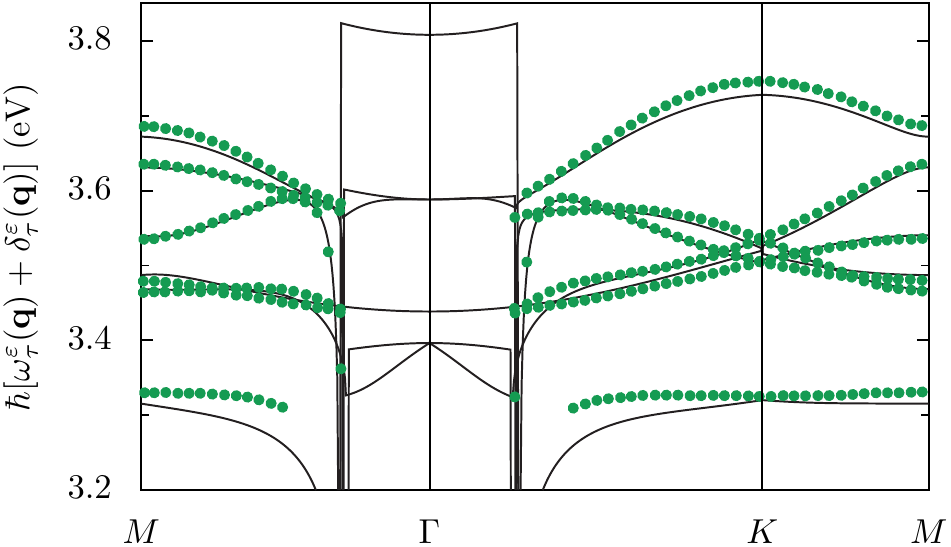}
\caption{\label{fig:Han} Retarded plasmonic dispersion relation of the honeycomb lattice for 
silver nanoparticles with resonance frequency $\omega_0=\unit[3.5]{eV/\hbar}$, radius $a=\unit[10]{nm}$, and interparticle distance $d=\unit[35]{nm}$. The green data points are extracted from Fig.~1(c) in Ref.~\cite{Han-PRL2009}. The solid lines correspond to the results of our open quantum system approach.}
\end{center}
\end{figure}

We reproduce in Fig.~\ref{fig:Han} the results of Ref.~\cite{Han-PRL2009} (green dots), which we compare to our analytical results based on the open quantum system approach (solid lines). Note that the method of Ref.~\cite{Han-PRL2009} 
cannot access the plasmonic modes inside of the light cone (i.e., for $|\q|\lesssim k_0$).
As can be seen from the figure, the quantitative agreement between both theories is rather good for collective modes outside of the light cone, thus validating our approximate treatment of the light--matter coupling, at least in this region of the 1BZ.  

We also compared our results to the fully retarded classical calculations by Zhen \textit{et al.}\ \cite{zhen08_PRB} performed on a square lattice of silver particles with radius $a=\unit[25]{nm}$ spaced by an interparticle distance $d=\unit[75]{nm}$. There, we obtain a qualitative agreement with the classical simulations. The lack of a quantitative agreement in this case can be understood by the rather large spacing between nanoparticles, for which $k_0d\simeq1.4$, which is clearly out of the range of validity of our approach ($k_0d\ll1$).

\bibliography{Biblio}

\end{document}